%% file: DistDetection_consensus_Part2_onecol.tex
\documentclass[journal,11pt,onecolumn]{IEEEtran}
    
\usepackage{epsf, psfrag, amssymb, amsfonts, amsmath, cite,enumerate}
\usepackage{graphicx, subfigure, color,bbm, bm}
\usepackage{boxedminipage}
\usepackage{multirow}
\usepackage{booktabs}
\usepackage{algorithm, algorithmic}
\usepackage{relsize}
\usepackage{exscale}
\input macro_jnl

\newtheorem{remark}{Remark}

\begin{document}
%
\title{Distributed Detection in Ad Hoc Networks Through Quantized Consensus} 
\author{Shengyu Zhu and Biao Chen
\thanks{This work was supported by the Air Force Office of Scientific Research under Award FA9550-16-1-0077. This paper was presented in part at the IEEE International Symposium on Information Theory (ISIT), Barcelona, Spain, July 2016 \cite{Zhu2016ISIT}.}
\thanks{
S.~Zhu and B.~Chen are with the Department of Electrical Engineering and Computer Science, Syracuse University, Syracuse, NY 13244, USA (e-mail: szhu05@syr.edu).}
}
\maketitle

\begin{abstract}
We study asymptotic performance of distributed detection in large scale connected sensor networks. Contrasting to the canonical parallel network where a single node has access to local decisions from all other nodes, each node can only exchange information with its direct neighbors in the present setting. We establish that, with each node employing an identical one-bit quantizer for local information exchange, a novel consensus reaching approach can achieve the optimal asymptotic performance of centralized detection as the network size scales. The statement is true under three different detection frameworks: the Bayesian criterion where the maximum {\em a posteriori} detector is optimal, the Neyman-Pearson criterion with a constant type-I error probability constraint, and the Neyman-Pearson criterion with an exponential type-I error probability constraint. Leveraging recent development in distributed consensus reaching using bounded quantizers with possibly unbounded data (which are log-likelihood ratios of local observations in the context of distributed detection), we design a one-bit deterministic quantizer with controllable threshold that leads to desirable consensus error bounds. The obtained bounds are key to establishing the optimal asymptotic detection performance.  In addition, we examine non-asymptotic performance of the proposed approach and show that the type-I and type-II error probabilities at each node can be made arbitrarily close to the centralized ones simultaneously when a continuity condition is satisfied. 
\end{abstract}
\begin{IEEEkeywords}
Distributed detection, error exponent, one-bit quantizer, quantized consensus, large deviations.
\end{IEEEkeywords}
\section{Introduction}
\IEEEPARstart{D}{istributed} detection in sensor networks has been an important research topic over the past decades \cite{Varshney1997distributed,Viswanathan1997,Blum1997,Chamberland2003,Chamberland2004,Tsitsiklis1988,Tay2008}. A canonical structure in distributed detection is the parallel fusion network where each sensor receives an observation about a common phenomenon and sends a local decision (e.g., the observation itself, the log-likelihood ratio (LLR), or its quantized version) to a fusion center. The fusion center makes the final decision based on all the information collected from the sensors. Another extensively studied structure is the tandem network in which local decisions propagate in a serial manner until they reach the last sensor that serves as the fusion center. Separately, there exist a large body of literature that deal with consensus type network inference problems in the absence of any fusion center (see, e.g., \cite{Saligrama2006,Kar2008,Kar2011,Kar2013,Jakovetic2012,Bajovic2012,Braca2010}). Sensors iteratively exchange information with their neighbors to arrive at a consensus decision using local updates. 

This paper investigates asymptotic performance of distributed detection in an $n$-node sensor network where each sensor receives an observation whose distribution is assumed to be  independent and identically distributed (i.i.d.) across all sensors. We specifically study the decay rate of detection error probability under the maximum \emph{a posteriori} (MAP) and the Neyman-Pearson criteria as the network size scales. The performance of distributed detection is clearly bounded by that of centralized detection where all the observations are available for decision making. With i.i.d.~observations, the optimal acceptance region in the centralized case is fully characterized by the average LLR of the observations. As such, the fusion or consensus based structures can achieve the optimal error exponent if sensors are able to communicate real values of infinite precision. Notice that the parallel and tandem networks respectively take $n$ and $n-1$ data transmissions, while the consensus based structure may require longer transmission time that depends on the consensus algorithm. 

Communicating real data, however, requires unlimited channel bandwidth in theory. Practical bandwidth and resource limitations often dictate that only quantized data can be reliably exchanged. Of particular interest is the extreme case where each sensor can only send one-bit information. Tsitsiklis established in \cite{Tsitsiklis1988} the optimality of identical likelihood ratio quantizers in such a setting for a canonical fusion network with communications allowed from the sensor to the fusion center (i.e., no consensus type iterations). With i.i.d.~observations across sensors, the decay rate under the Neyman-Pearson criterion is determined by the Kullback-Leibler divergence of the binary output and is typically smaller than the centralized one. Under the MAP criterion, the decay rate is also suboptimal to the centralized one in general. For the tandem network, it was shown in \cite{Tay2008} that using a one-bit quantizer at each sensor can never achieve an exponential decay rate of the error probability under the MAP criterion. To the best of our knowledge, there is no asymptotic result on consensus based structures using one-bit quantizers at each node. As such, it is {\emph{a priori}} unknown whether one-bit quantization in a general connected network that allows iterative message exchanges can achieve exponentially decaying error probability and what would be the optimal exponent if exponentially vanishing error probability is feasible. Note that if nodes have perfect knowledge of global network topology, one may construct schemes that utilize source coding ideas to attain the same optimal error exponent as in the centralized setting. This, however, is not realistic in most applications where nodes only have knowledge of their directly connected neighbors.

In this paper, we consider distributed detection over general connected sensor networks using iterative distributed averaging algorithms, with the goal of reaching a consensus on the average LLR among all sensors in the network. Sensors can only exchange one-bit information with their immediate neighbors and have no knowledge about the global topology except the fact that the network is connected. Both parallel and tandem networks can be included in this setup if sensors and fusion center are able to transmit and receive data. Our motivation is to broaden the appeal of decentralized inference and to examine problems that are relevant to many emerging applications involving large scale networks of arbitrary topology. Before stating our approach, we would like to briefly review existing distributed averaging algorithms with quantized communications to illustrate why there are no asymptotic results on consensus based detection using one-bit communications:
\begin{enumerate}[a)]
\item most quantized consensus algorithms use infinite-level quantizers (e.g., rounding quantizer and truncated quantizer) which still require unlimited bandwidth; see \cite{Kashyap2007,Nedic2009,Kar2010,Carli2010,Zhu2016TSP}, among others.
\item while some recent results \cite{Kar2010,Li2011} can work with finite-bit quantizers, they do not guarantee convergence at a consensus in finite time. More critically, they assume that the data for average consensus are bounded and the bound is known {\emph{a priori}}.{\footnote{Different from the algorithms of part a) where truncation may be used to get a finite-bit quantizer when knowing the bound, the algorithms in \cite{Kar2010,Li2011} use the bound to choose appropriate algorithm parameters for predefined finite-bit quantizers.}} This assumption is generally too restrictive for detection since the LLRs of local observations can be arbitrarily large for common distributions, e.g., Gaussian distributions with different means for the two hypotheses.  Specifically, the design and analysis of distributed consensus with finite (bounded) quantizer and unbounded input appear to be key in solving the detection problem at hand.
\item a novel distributed averaging algorithm, referred to as BQ-CADMM, is proposed in  \cite{Zhu2016BQC}. The algorithm uses a finite-bit bounded quantizer  which first projects its argument to a compact set and then applies rounding quantizer to the projected value. Within finite iterations, the quantized variable values at all nodes either converge to the same quantization level or cycle around the average with the same sample mean over a finite period. The consensus value is subject to a consensus error from the desired average due to quantized communications. Though an upper bound is obtained for the consensus error, it has a non-vanishing constant error term and the resulting consensus cannot be arbitrarily close to the true average in general. 
\end{enumerate}

Despite the presence of consensus errors, we note that BQ-CADMM has the advantages that it achieves a consensus within finite iterations and that it does not assume any bound on the input data. Additionally, the constant term in the error bound is from the quantization error of the rounding quantizer operating on the projected value. {\emph{We therefore adopt a new one-bit quantizer with controllable threshold such that the consensus error can be made arbitrarily small if the consensus is reached at a specific quantization point}. With this quantizer, we proceed to construct an acceptance region using the consensus result and show that it approaches the optimal acceptance region in the centralized case asymptotically. This helps establish that the best achievable rate is the same as the centralized one under the MAP and the Neyman-Pearson criteria. Non-asymptotic scenarios are also studied and the error probabilities at each node can be made arbitrarily close to the optimal centralized ones under a continuity condition.


The rest of the paper is organized as follows. Section~\ref{sec:Pre} reviews useful concepts and results. Section \ref{sec:BQC} defines a new one-bit quantizer and establishes the convergence result of BQ-CADMM with this quantizer. The quantizer is tailored for the detection application such that desired consensus accuracy is guaranteed. In Section \ref{sec:result}, the consensus result of BQ-CADMM is used to construct an acceptance region that is shown to approach the optimal centralized one and hence to achieve the same error exponent under the MAP and Neyman-Pearson criteria. Section~\ref{sec:nonasymp} studies the non-asymptotic performance of the proposed approach. Simulations are provided in Section~\ref{sec:simulation} where numerical examples are used to evaluate the proposed approach and to characterize its convergence time. Section~\ref{sec:conclusion} concludes the paper and discusses further research directions.
\section{Problem and Preliminary}
\label{sec:Pre}
Section~\ref{sec:ProbState} states the problem of this paper and Section~\ref{sec:Pre2} introduces preliminary concepts and results that will be used to prove our main results in Section~\ref{sec:result}.
\subsection{Problem Statement}
\label{sec:ProbState}
Consider a connected $n$-node sensor network with $m$ bi-directional links. Each sensor $i,i=1,2,\ldots,n$, has its own observation $y_i$. We model this network as a connected undirected graph with $n$ nodes and $m$ edges. Let $\mathcal{N}_i$ denote the set of directly linked nodes of node $i$ and $|\mathcal{N}_i|$ its cardinality. {Here node $i$ is not considered as a linked node of itself, i.e., $i\notin\mathcal{N}_i$.} Then $n-1\leq m\leq\frac{n(n-1)}{2}$, $1\leq|\mathcal{N}_i|\leq n-1$, and $2m =\sum_{i=1}^n|\mathcal{N}_i|$. Assume that the observations $y^n$ originate from an i.i.d.~source $Q(y)$ with alphabet $\Sigma$ that can be either a finite set or any Polish space.\footnote{That is, separable completely metrizable topological space. We follow \cite{Dembo2009} to use Polish space in order not to be distracted by measurability concerns. In many applications like ours, $\Sigma$ is either a finite set or a subset of $\mathbb{R}^d$ for some $d\in\mathbb{Z}^+$.} Denote $\mathcal{P}(\Sigma)$ as the space of probability measures on $\Sigma$. We consider two hypotheses
\begin{itemize}
\item $H_1:Q=P_1\in\mathcal{P}(\Sigma),$
\item $H_2:Q=P_2\in\mathcal{P}(\Sigma),$
\end{itemize}
with prior probabilities $\pi_1$ and $\pi_2=1-\pi_1$, respectively. Let $\mathcal{A}_n\subseteq\Sigma^n$ denote the acceptance region for $H_1$ and $\mathcal{A}_n^c=\Sigma^n\setminus\mathcal{A}_n$ the critical region. Then the {\it type-I and type-II error probabilities} are respectively $$\alpha_n = P_1(\mathcal{A}_n^c)~\text{and}~\beta_n = P_2(\mathcal{A}_n).$$

We investigate the asymptotic detection performance via consensus based approaches where sensors can only reliably exchange one-bit information with its neighbors at each iteration. To ensure autonomy in a large sensor network, local sensors, or more precisely, local computations and communications, do not require the information about the global network structure. We consider the following three criteria for large connected sensor networks:
\begin{itemize}
\item Neyman-Pearson criterion with constant constraint: for a given $\alpha\in(0,1)$,
\begin{equation}
\begin{aligned}
&\text{maximize} &&\liminf_{n\to\infty}-\frac{1}{n}\log \beta_n,\\
&\text{subject to}&&\lim_{n\to\infty}\alpha_n\leq\alpha.\nn
\end{aligned}
\end{equation}
\item MAP criterion: given $\pi_1, \pi_2\in(0,1)$,
\begin{align}
\text{maximize}~\liminf_{n\to\infty}-\frac{1}{n}\log\left(\pi_1\alpha_n+\pi_2\beta_n\right).\nn
\end{align}
\item Neyman-Pearson criterion with exponential constraint: for a given $\gamma\in(0,D(P_2||P_1))$,
\begin{equation}
\begin{aligned}
&\text{maximize}&&\liminf_{n\to\infty}-\frac{1}{n}\log\beta_n,\\~&\text{subject to}&& \liminf_{n\to\infty}-\frac1n\log\alpha_n\geq\gamma.\nn
\end{aligned}
\end{equation}
\end{itemize}

It is {\em a priori} unknown a) if exponentially vanishing error probabilities in the network size can be achieved given only one-bit local information exchange; and b) what would be the optimal error exponent if indeed exponentially decaying error probabilities can be attained. Before introducing the consensus based scheme for the construction of acceptance regions, we first review centralized results that act as performance bounds for distributed detection.

\subsection{Preliminaries}
\label{sec:Pre2}
Throughout the rest of this paper, we assume that $P_1\in\mathcal{P}(\Sigma)$ and $P_2\in\mathcal{P}(\Sigma)$ are absolutely mutually continuous. We begin with the definition of relative entropy.
\begin{definition}
The relative entropy or Kullback-Leibler divergence between $P_1$ and $P_2$ is defined as 
\begin{align}
D(P_1\|P_2) =\int_{\Sigma} \log\frac{dP_1}{dP_2}dP_1=\mathbb{E}_{P_1}\left(\log\frac{dP_1}{dP_2}\right),\nn
\end{align}
where $dP_1/dP_2$ stands for the Radon-Nikodym derivative of $P_1$ with respect to $P_2$.
\end{definition}

For ease of presentation, we write $dP_1$ and $dP_2$ as $p_1$ and $p_2$, respectively. Using the weak law of large numbers, we can derive the following asymptotic equipartition property for the relative entropy.

\begin{theorem}[{\hspace{-0.2pt}\cite[Theorem 11.8.1]{Cover2006}}]
\label{thm:aepr}
Let $y^n$ be a sequence of random variables drawn i.i.d.~according to $P_1$, and let $P_2$ be any other measure from $\mathcal{P}(\Sigma)$. Then
\begin{align}
\frac{1}{n}\log\frac{p_1(y^n)}{p_2(y^n)}\rightarrow D(P_1\|P_2)~\text{in probability}.\nn
\end{align}
\end{theorem}

\begin{definition}
For a fixed $n$ and $\epsilon> 0$, a sequence $y^n\in\Sigma^n$ is said to be relative entropy typical if and only if
$$D(P_1\|P_2)-\epsilon\leq\frac{1}{n}\log\frac{p_1(y^n)}{p_2(y^n)}\leq D(P_1\|P_2)+\epsilon.$$
The set of relative entropy typical sequences is called the relative entropy
typical set $A_\epsilon^{(n)}(P_1\|P_2)$.
\end{definition}

We then have the following lemma as a direct consequence of Theorem~\ref{thm:aepr}. 
\begin{lemma}[\hspace{-0.2pt}{\cite[Theorem 11.8.2]{Cover2006}}]
\label{lem:pro}
Given any positive $\epsilon$, 
$$P_1\left(A_\epsilon^{(n)}(P_1\|P_2)\right) > 1-\epsilon,$$
provided that $n$ is sufficiently large.
\end{lemma}

With the above definitions, we are ready to present Stein's lemma which provides the best exponent for the type-II error probability under the Neyman-Pearson criterion with a constant constraint on the type-I error probability.
\begin{theorem}[Stein's Lemma{\cite[Theorem 11.8.3]{Cover2006}, \cite[Lemma 3.4.7]{Dembo2009}}]
Let $y^n$ be i.i.d.~$\sim Q$. Consider the test between two hypotheses $H_1:Q=P_1$ and $H_2:Q=P_2$, where $0<D(P_1\|P_2) < \infty$. Let $\beta_n^\alpha$ be the infimum of $\beta_n$ among all
tests with $\alpha_n\leq\alpha$. Then for any $0<\alpha<1$, 
$$\lim_{n\to\infty}-\frac{1}{n}\log\beta_n^\alpha=D(P_1\|P_2), $$ which can be asymptotically achieved by choosing the acceptance region as $A_\epsilon^{(n)}(P_1\|P_2)$ with $\epsilon \to 0$, i.e., 
$$\lim_{\epsilon\to0}\lim_{n\to\infty}-\frac{1}{n}\log P_2\left(A_\epsilon^{(n)}(P_1\|P_2)\right)=D(P_1\|P_2).$$ 
\end{theorem}

Under the Bayesian framework, the best error exponent is provided by Chernoff theorem.
\begin{theorem}[Chernoff{\cite[Theorem 11.9.1]{Cover2006}, \cite[Lemma 3.4.6]{Dembo2009}}]
For $\pi_1>0$ and $\pi_2>0$, the best achievable exponent in the Bayesian probability of error is given by
$$\liminf_{n\to\infty}-\frac{1}{n}\log(\pi_1\alpha_n+\pi_2\beta_n)=C(P_1,P_2),$$ where $C(P_1,P_2)$ is the Chernoff information defined as
$$C(P_1,P_2)\triangleq-\underset{0\leq\lambda\leq 1}{\min}\log\left(\int_\Sigma p_1(y)^\lambda p_2(y)^{1-\lambda}dy\right).$$
\end{theorem}

We next present the centralized result under the Neyman-Pearson criterion with exponential constraint via large deviations. Define the logarithmic moment generating function of the LLR as
$$\Lambda(\lambda) = \log\mathbb{E}_{P_1}\left(e^{-\lambda \log \frac{p_1(y)}{p_2(y)}}\right),\lambda\in\mathbb{R}.$$
Notice that $\Lambda(0)=\Lambda(1)=0$ for the hypothesis testing problem as $P_1$ and $P_2$ are assumed to be mutually absolutely continuous. The Fenchel-Legendre transform of $\Lambda(\lambda)$, which characterizes the large deviations associated with the empirical mean of i.i.d. random variables, is defined as
$$\Lambda^*(\tau) \triangleq \sup_{\lambda\in\mathbb{R}}\{\lambda\tau - \Lambda(\lambda)\}.$$
A useful property of $\Lambda^*(\cdot)$ is stated in the following lemma, which is a direct result from \cite[Lemma~2.2.5]{Dembo2009}.
\begin{lemma}
\label{lem:Lambda}
$\Lambda^*(\tau)$ is a non-decreasing convex function for $\tau> -D(P_1\|P_2)$.
\end{lemma}

The following theorem characterizes the large deviations of the probabilities of error under likelihood ratio tests.
\begin{theorem}[\hspace{-0.1pt}{\cite[Theorem~3.4.3]{Dembo2009}}]
\label{thm:largedev}
Let the acceptance region for $H_1$ be 
\begin{align}
\label{eqn:NPtest}
\left\{y^n:\frac{1}{n}\log\frac{p_1(y^n)}{p_2(y^n)}>-\tau\right\}. 
\end{align} 
Given $\tau\in(-D(P_1\|P_2), D(P_2\|P_1))$, the error probabilities satisfy
\begin{align}
\lim_{n\to\infty}-\frac{1}{n}\log\alpha_n=\Lambda^*(\tau)>0,\nn
\end{align}
and
\begin{align}
\lim_{n\to\infty}-\frac{1}{n}\log\beta_n=\Lambda^*(\tau)-\tau>0.\nn
\end{align}
\end{theorem}

The acceptance region in (\ref{eqn:NPtest}) is referred to as the Neyman-Pearson test in the literature. It is straightforward to see that the type-II error probability $\beta_n$ becomes larger as $\tau$ increases. Thus, Theorem~\ref{thm:largedev} together with the optimality of the Neyman-Pearson test (see, e.g., \cite{Cover2006,Dembo2009}) implies that the optimal error exponent under the Neyman-Pearson criterion with exponential constraint is given by $$\liminf_{n\to\infty}-\frac{1}{n}\log\beta_n=\Lambda^*(\tau^*)-\tau^*,$$
where $\tau^*$ is the smallest value in $(-D(P_1\|P_2),D(P_2\|P_1))$ such that $\Lambda^*(\tau^*)=\gamma$ and its existence is guaranteed as per Stein's lemma and Lemma~\ref{lem:Lambda}. The corresponding acceptance region is then given by the Neyman-Pearson test in (\ref{eqn:NPtest}) with $\tau=\tau^*$. 

Also shown in \cite{Dembo2009}, both Stein's lemma and Chernoff theorem can be deduced from Theorem~\ref{thm:largedev}. An interesting fact is that the Chernoff information is equal to the Fenchel-Legendre transform of $\Lambda(\cdot)$ evaluated at zero, i.e., $C(P_1,P_2)=\Lambda^*(0)$. Instead of directly studying the Neyman-Pearson criterion with exponential constraint via large deviations, we will first consider the Neyman-Pearson criterion with constant constraint and the Bayesian criterion in Section~\ref{sec:result} to help illustrate our approach. In order to apply consensus based approaches, notice that for all the three criteria reviewed in Section~\ref{sec:Pre2}, the optimal detectors all amount to a form of threshold test of the global LLR. With distributed detection, if one can reconstruct the global LLR, then optimal detection performance in the centralized setting can be attained. Since the global LLR is equivalent to the average of all local LLR values, this motivates the average consensus approach for distributed detection where local LLRs are treated as local agent data. We comment again that the LLRs for most observation models are intrinsically unbounded. The next section introduces such a distributed averaging algorithm that uses only one-bit quantizer at each node.
\section{Distributed Average Consensus using One-Bit Communications}
\label{sec:BQC}

The BQ-CADMM approach proposed in \cite{Zhu2016BQC} employs a finite-bit quantizer that applies a projection operator followed by the uniform rounding quantizer. Due to this rounding quantizer, the resulting consensus value is subject to a consensus error from the desired average and the derived error bound has a non-vanishing constant error term (cf. \cite[Theorem~3]{Zhu2016BQC}). This fact implies that the consensus based approach is in itself insufficient if one is to attain the same asymptotic performance of the centralized case. We therefore adopt a new binary quantizer with controllable threshold such that the consensus error can be made arbitrarily small when consensus is reached at a specific quantization point. 

We now construct the one-bit quantizer in a similar fashion to that of \cite{Zhu2016BQC}: the composition of (uniform) quantization and projection. Given quantization resolution $\Delta>0$ and a predefined quantization point $a\in\mathbb{R}$, let $\mathcal{Q}(\cdot)$ be a uniform quantizer defined as 
$$\mathcal{Q}(x) = a+t\Delta,~\text{if}~a+t\Delta-\delta <x\leq a+(t+1)\Delta-\delta,$$ where $x\in\mathbb{R}$, $t\in\mathbb{Z}$, and $\delta\in(0,\Delta)$. If we pick $a=0$ and $\delta=\frac{\Delta}{2}$, then $\mathcal{Q}(\cdot)$ becomes the usual rounding quantizer. Let $\mathcal{X}=[a,a+\Delta]$ and denote by $\mathcal{T_X}:\mathbb{R}\to\mathcal{X}$ the projection operator that maps $x\in\mathbb{R}$ to the nearest point in $\mathcal{X}$, i.e.,
\begin{align}
\mathcal{T_X}(x)=\begin{cases}
a,&~\text{if}~x<a,\\
x,&~\text{if}~a\leq x\leq a+\Delta,\\
a+\Delta,&~\text{otherwise}.\nn
\end{cases}
\end{align}
The one-bit quantizer is defined as
\begin{align}
\label{eqn:detalQDef}
\mathcal{Q}_\delta(\cdot)=\mathcal{Q}\circ\mathcal{T_X}(\cdot),
\end{align}
which we refer to as $\delta$-quantizer. One can easily verify that the $\delta$-quantizer is equivalent to a binary threshold quantizer
\begin{align}
\mathcal{Q}_\delta(x)=\begin{cases}
a,&~\text{if}~x\leq a+\Delta-\delta,\\
a+\Delta,&~\text{otherwise}.\nn
\end{cases}
\end{align}

Presented in Algorithm \ref{tab:QCADMM} is BQ-CADMM with this $\delta$-quantizer, where $r_i\in\mathbb{R}$ denotes the local data, i.e., local LLR at node $i$, and $\rho$ is the algorithm parameter that can be any positive value. It is straightforward to see that BQ-CADMM is fully distributed since the updates of local variables $x_i^{k+1}$ and $\alpha_i^{k+1}$ only rely on local and neighboring information.{\footnote{Throughout the rest of this paper, `BQ-CADMM' stands for the algorithm with the $\delta$-quantizer; we use `original BQ-CADMM' to represent the algorithm with the bounded rounding quantizer in \cite{Zhu2016BQC}.}} While similar results of BQ-CADMM using this $\delta$-quantizer can be obtained as a direct generalization of \cite{Zhu2016BQC}, we derive tighter consensus error bounds  by taking advantage of  the fact that there are only two quantization values. This is stated in Thereom~\ref{thm:BQCADMMdelta}.

\begin{center}
\begin{minipage}{0.8\linewidth}
\begin{algorithm}[H]
	\caption{BQ-CADMM with the $\delta$-quantizer}
	\begin{algorithmic}[1]
	\label{tab:QCADMM}
	\REQUIRE Initialize~$x_i^0=0$ and $\alpha_{i}^0=0$ for each agent $i,i=1,2,\ldots,n$. Set $\rho>0$ and $k=0$.
	\REPEAT
			\STATE every agent $i$ {\bf do}
			\begin{equation} 
			\begin{aligned}
			x_i^{k+1}=&~\frac{1}{1+2\rho|\mathcal{N}_i|}\Bigg(\rho|\mathcal{N}_i|\mathcal{Q}_\delta(x_i^k)+\rho\sum_{j\in\mathcal{N}_i}\mathcal{Q}_\delta(x_j^k)-\alpha_i^k+r_i\Bigg),\nn\\
\alpha_i^{k+1}=&~\alpha_i^k+\rho\Bigg(|\mathcal{N}_i|\mathcal{Q}_\delta(x_{i}^{k+1})-\sum_{j\in\mathcal{N}_i}\mathcal{Q}_\delta(x_{j}^{k+1})\Bigg).
			\end{aligned}
			\end{equation}
			\STATE {\bf set} $k=k+1$.	
	\UNTIL{a predefined stopping criterion (e.g., a maximum iteration number) is satisfied.}
	\end{algorithmic}
\end{algorithm}
\end{minipage}
\end{center}
\hspace{1em}
\begin{theorem}
\label{thm:BQCADMMdelta}
Let $\bar{r}=\frac{1}{n}\sum_{i=1}^nr_i$ denote the data average. For BQ-CADMM using the $\delta$-quantizer $\mathcal{Q}_\delta(\cdot)$, there exists a finite number of iterations $k_0$ such that for $k\geq k_0$, all the quantized variable values
\begin{itemize}\setlength{\leftmargin}{0pt}
\item either converge to the same quantization value:
 $$\mathcal{Q}_\delta(x_1^k)=\mathcal{Q}_\delta(x_2^k)=\cdots=\mathcal{Q}_\delta(x_n^k)\triangleq x_Q^*\in\{a,a+\Delta\},$$ where $x_Q^*$ satisfies the following error bound
\begin{align}
\label{eqn:bounddQ}
\begin{cases}\left|x_Q^*\hspace{-1pt}-\hspace{-1pt}\mathcal{T_X}(\bar{r})\right|\leq
\left(1+4\rho\frac{m}{n}\right)(\Delta\hspace{-1pt}-\hspace{-1pt}\delta),&\text{if}~x_Q^*=a,\\
\left|x_Q^*\hspace{-1pt}-\hspace{-1pt}\mathcal{T_X}(\bar{r})\right|<\left(1+4\rho\frac{m}{n}\right)\delta,&\text{if}~x_Q^*=a+\Delta,
\end{cases}
\end{align}
\item or cycle around the true average $\bar{r}$ with a finite period $T\geq2$, i.e.,
$ x_i^k=x_i^{k+T},i=1,2,\ldots,n$. Furthermore, \begin{align}
\label{eqn:cyceqn}
\sum_{l=0}^{T-1}\mathcal{Q}_\delta(x_1^{k+l})=\sum_{l=0}^{T-1}\mathcal{Q}_\delta(x_2^{k+l})=\cdots=\sum_{l=0}^{T-1}\mathcal{Q}_\delta(x_n^{k+l}),\end{align}
and 
\begin{align}
\label{eqn:boundcyclic}
\left|\bar{r}-(a+\Delta-\delta)\right|<6\rho n\Delta.
\end{align}
\end{itemize}

\end{theorem}
\begin{IEEEproof}
See Appendix.
\end{IEEEproof}
\begin{remark}
Contrasting with the original BQ-CADMM that has a uniform consensus error bound when the algorithm converges, the $\delta$-quantizer results in error bounds that are dependent on the consensus value. This is achieved by the asymmetric rounding of the $\delta$-quantizer. Clearly, choosing a small $\delta$ relative to $\frac{\Delta}{2}$ will skew the quantizer toward $a$, i.e., the quantizer threshold is much closer to $a+\Delta$. Thus, when consensus is reached at $a+\Delta$, the consensus error is ensured to be small too.
\end{remark}
\begin{remark}
While Theorem~\ref{thm:BQCADMMdelta} only requires a connected network, the convergence time (the smallest $k_0$ in convergent cases or the smallest $k_0+T$ in cyclic cases) depends on the agents' data, the network structure as well as the algorithm parameter $\rho$. Besides, BQ-CADMM converges in most cases, particularly with large and dense networks or small enough algorithm parameters (see simulations in Section~\ref{sec:simulation}). 
\end{remark}
{
\begin{remark}
For BQ-CADMM to work, i.e., Theorem~\ref{thm:BQCADMMdelta} to hold, the algorithm parameter $\rho$ can be any positive value and does not depend on other parameters; however, to guarantee certain accuracy as we will need in the next section, $\rho$ has to be selected according to network parameters such as the number of nodes and the number of edges. In addition, the choice of $\rho$ has an impact on whether convergence or oscillation can happen. To see this, consider $\bar{r}\neq a+\Delta-\delta$. Then (\ref{eqn:boundcyclic}) is violated with small enough $\rho$ and convergence must be reached. This might  also explain why small algorithm parameters are likely to yield convergence results in our simulations where $\bar{r}$ itself is random.
\end{remark}}

\section{Optimal Asymptotic Performance}
\label{sec:result}
This section establishes the optimal asymptotic performance under the three criteria. We  use the consensus result from BQ-CADMM with appropriate algorithm parameter and quantizer setup to construct acceptance regions that can asymptotically achieve the optimal performance in centralized settings. 

\subsection{Neyman-Pearson Criterion with Constant Constraint}
\label{sec:NPC}
From Stein's lemma, the relative entropy typical set $A_\epsilon^{(n)}(P_1\|P_2)$ achieves the optimal error exponent in the centralized setting with diminishing $\epsilon$. Consequently, by picking suitable $\rho$ and small enough $\delta$ we can construct an acceptance region that is asymptotically equivalent to ${A}_\epsilon^{(n)}(P_1\|P_2)$ thus achieving the same error exponent. The result is presented in the following theorem.
\begin{theorem}
\label{thm:mainresult}
Let $y^n$ be i.i.d.~$\sim Q$. Consider the test between two hypotheses $H_1:Q=P_1$ and $H_2:Q=P_2$, where $0<D(P_1\|P_2) < \infty$. Let $r_i=\log\frac{p_1(y_i)}{p_2(y_i)}$ be the local data at node $i$. Set $a=0$, $\Delta=D(P_1\|P_2)$, and $$\rho=\min\left\{\frac{\delta}{6nD(P_1\|P_2)},\frac{n}{4m}\right\}.$$ Assume that BQ-CADMM runs sufficiently long such that either convergence or cycling occurs. Let the acceptance region for $H_1$ be
\begin{align}
\mathcal{A}_n^\delta=\big\{y^n:\text{BQ-CADMM cycles}\}\hspace{2pt}\mathsmaller{\bigcup}\hspace{2pt}\big\{y^n:\text{BQ-CADMM converges at~}x_Q^*=D(P_1\|P_2)\big\}\nn.
\end{align}
Then given any $0<\delta<D(P_1\|P_2)$,
\begin{align}
\label{eqn:type1}
P_1\left(\left(\mathcal{A}_n^\delta\right)^c\right)<\frac{1}{2}\delta,~\text{for}~n~\text{sufficiently large},
\end{align}
and moreover,
\begin{align}
\label{eqn:type2}
\lim_{\delta\to0}\lim_{n\to\infty}-\frac{1}{n}\log P_2\left(\mathcal{A}_n^\delta\right)=D(P_1\|P_2).
\end{align}
\end{theorem}
\begin{IEEEproof} In this case we have $\bar{r}=\frac{1}{n}\log\frac{p_1(y^n)}{p_2(y^n)}$, the average LLR. We first find a {\emph{sufficient}} condition for $y^n\in\mathcal{A}_n^\delta$ to establish (\ref{eqn:type1}). If $y^n\notin\mathcal{A}_n^\delta$, then BQ-CADMM converges at $x_Q^*=0$ and hence (\ref{eqn:bounddQ}) implies  $$\left|\mathcal{T_X}\left(\frac{1}{n}\log\frac{p_1(y^n)}{p_2(y^n)}\right)\right|\leq\left(1+4\rho\frac{m}{n}\right)(D(P_1\|P_2)-\delta).$$
Picking $0<\rho\leq\rho_1\triangleq\frac{n\delta}{9m(D(P_1\|P_2)-\delta)}$, we have $$\left(1+4\rho\frac{m}{n}\right)(D(P_1\|P_2)-\delta)\leq D(P_1\|P_2)-\frac{5}{9}\delta.$$ Thus, if $y^n$ is such that $\left|\frac{1}{n}\log\frac{p_1(y^n)}{p_2(y^n)}-D(P_1\|P_2)\right|\leq\frac{1}{2}\delta$, $y^n$ must lie in $\mathcal{A}_n^\delta$ as $\mathcal{T_X}(\cdot)$ projects a real value to the nearest point in $\mathcal{X}=[0,D(P_1\|P_2)]$. Therefore, 
\begin{align}
\mathcal{A}_n^{\delta}&\supseteq\left\{y^n:\left|\frac{1}{n}\log\frac{p_1(y^n)}{p_2(y^n)}-D(P_1\|P_2)\right|\leq\frac{1}{2}\delta\right\}=A^{(n)}_{\delta/2}(P_1\|P_2).\nn
\end{align}
Hence, (\ref{eqn:type1}) is true according to Lemma \ref{lem:pro}.

We next show that $\mathcal{A}_n^\delta$ can asymptotically achieve the optimal error exponent by identifying a \emph{necessary} condition for $y^n\in\mathcal{A}_n^\delta$. When convergence happens, $x_Q^*=D(P_1\|P_2)$ and (\ref{eqn:bounddQ}) implies that
\begin{align}
\left|\mathcal{T_X}\left(\frac{1}{n}\log\frac{p_1(y^n)}{p_2(y^n)}\right)-D(P_1\|P_2)\right|&<\left(1+4\rho\frac{m}{n}\right)\delta.\nn
\end{align}
If we again pick $\rho$ small enough, e.g., $0<\rho\leq\rho_2\triangleq\frac{n}{4m}$, then $\left(1+4\rho\frac{m}{n}\right)\delta\leq2\delta$. Thus, $$\mathcal{T_X}\left(\frac{1}{n}\log\frac{p_1(y^n)}{p_2(y^n)}\right)> D(P_1\|P_2)-2\delta.$$
Assume that $0<\delta<\frac{D(P_1\|P_2)}{2}$ such that $D(P_1\|P_2)-2\delta>0$. From the definition of $\mathcal{T_X}(\cdot)$, we have 
\begin{align}
\label{eqn:convergenceAnd}
\frac{1}{n}\log\frac{p_1(y^n)}{p_2(y^n)}> D(P_1\|P_2)-2\delta.
\end{align}
Now if BQ-CADMM cycles, we have from Theorem \ref{thm:BQCADMMdelta} that 
\begin{align}
\left|\frac{1}{n}\log\frac{p_1(y^n)}{p_2(y^n)}-\left(D(P_1\|P_2)-\delta\right)\right|<6\rho n\Delta.\nn
\end{align}
Letting $\rho\leq\rho_3=\frac{\delta}{6nD(P_1\|P_2)}$, we conclude that (\ref{eqn:convergenceAnd}) is also true. Thus, if $y^n\in\mathcal{A}_n^\delta$ and $0<\delta<\frac{D(P_1\|P_2)}{2}$, we have
$$p_2(y^n)< p_1(y^n)2^{-n(D(P_1\|P_2)-2\delta)},$$
and further,
\begin{align}
-\frac{1}{n}\log P_2(\mathcal{A}_n^\delta)&=-\frac{1}{n}\log\int_{\mathcal{A}_n^\delta}p_2(y^n)dy\nn\\
&>-\frac{1}{n}\log\left(2^{-n(D(P_1\|P_2)-2\delta)}\int_{\mathcal{A}_n^\delta}p_1(y^n)dy\right)\nn\\
&\geq {D(P_1\|P_2)}-2\delta,\nn
\end{align}
which, together with Stein's lemma, implies (\ref{eqn:type2}).

The proof is complete by choosing $\rho=\min\{\rho_1,\rho_2,\rho_3\}$ and noting that $\delta<D(P_1\|P_2)$ and $m\leq\frac{n(n-1)}{2}$. 
\end{IEEEproof}

Therefore, by choosing small enough $\delta$, we have that $\alpha_n<\frac{1}{2}\delta\leq \alpha$ for large $n$ and that the type-II error exponent is arbitrarily close to the optimal error exponent $D(P_1\|P_2)$ which is given in Stein's lemma. Moreover, the above proof implies that as long as $\delta\to0$ with $n\to\infty$, we can get 
\begin{align}
\label{eqn:onelimit}
\lim_{n\to\infty}-\frac{1}{n}\log \beta_n = D(P_1\|P_2).
\end{align}
On the other hand, $\delta$ cannot decrease too fast in order to satisfy the type-I error constraint. With finite alphabet, a simple choice to meet the type-I error constraint is from Hoeffding's test \cite{Hoeffding1965} by setting $\delta=\frac{|\Sigma|\log n}{n}$ where $|\Sigma|$ denotes the cardinality of the alphabet. In general, there is not such a universal selection  for $\delta$. We may calculate a $\delta>0$ such that $P_1\left(A_{\delta/2}^{(n)}\right)\geq 1-\alpha$, and it is not hard to show that this $\delta$ vanishes as $n$ scales. If the above choice is greater than or equal to $D(P_1\|P_2)$, we can simply set $\delta=\frac{D(P_1\|P_2)}{2}$ to ensure $0<\delta<D(P_1\|P_2)$. In this way, (\ref{eqn:onelimit}) is guaranteed under the type-I error constraint. 

\subsection{MAP Criterion}
\label{sec:MAP}
Unlike the Neyman-Pearson criterion with constant constraint, the MAP criterion does not require a diminishing $\delta$. As one will see, this is because the optimal acceptance region converges to the same set asymptotically for any positive prior probabilities. Specifically, it is well-known that the optimal acceptance region for any $n$ under the MAP criterion is 
\begin{align}
\label{eqn:MAPoptset}
\left\{y^n:\frac{1}{n}\log\frac{p_1(y^n)}{p_2(y^n)}> \frac{1}{n}\log\frac{\pi_2}{\pi_1}\right\},
\end{align}
provided that $\pi_1$ and $\pi_2$ are both positive. We also have
\begin{align}
\label{eqn:MAPoptset2}
&~\liminf_{n\to\infty}-\frac{1}{n}\log P_1\left(\left\{y^n:\frac{1}{n}\log\frac{p_1(y^n)}{p_2(y^n)}\leq\frac{1}{n}\log\frac{\pi_2}{\pi_1}\right\}\right)\nn\\=&~\liminf_{n\to\infty}-\frac{1}{n}\log P_2\left(\left\{y^n:\frac{1}{n}\log\frac{p_1(y^n)}{p_2(y^n)}> \frac{1}{n}\log\frac{\pi_2}{\pi_1}\right\}\right)\nn\\
=&~C(P_1,P_2),
\end{align}
as a result of Lemma~\ref{lem:Lambda}, Theorem~\ref{thm:largedev}, and the fact that $\Lambda^*(0)=C(P_1,P_2)$. We remark that (\ref{eqn:MAPoptset2}) does not depend on particular values of $\pi_1$ and $\pi_2$ as long as they are positive. The following theorem states our result on the exponent of the Bayesian error probability. 

\begin{theorem}
\label{thm:MAP}
Let $y^n$ be i.i.d~$\sim Q$. Consider the hypothesis test between $H_1:Q=P_1$ and $H_2:Q=P_2$ with positive prior probabilities $\pi_1$ and $\pi_2$, respectively. Assume that $C(P_1,P_2)>0$. For the $\delta$-quantizer $\mathcal{Q}_\delta(\cdot)$, set $a=-1$, $\Delta=2$, and $\delta =1$. Set also the local data $r_i=\log\frac{p_1(y_i)}{p_2(y_i)}$ and the algorithm parameter $\rho=\frac{1}{12n^2}$.  Assume that BQ-CADMM runs sufficiently long such that either convergence or cycling occurs. Let the acceptance region for $H_1$ be
\begin{align}
\mathcal{A}_n=&\big\{y^n\hspace{-1pt}:\text{BQ-CADMM cycles}\}\hspace{2pt}\mathsmaller{\bigcup}\hspace{2pt}\big\{y^n\hspace{-1pt}:\text{BQ-CADMM converges at~}x_Q^*\hspace{-1pt}=1\big\}\hspace{-1pt}.\nn
\end{align}
Then the error exponent is given by$$\liminf_{n\to\infty}-\frac{1}{n}\log(\pi_1\alpha_n+\pi_2\beta_n)= C(P_1,P_2).$$
\end{theorem}
\begin{IEEEproof} With this setup, the $\delta$-quantizer has the threshold at $a+\Delta-\delta=0$. First note that if BQ-CADMM results in oscillation between the two quantization points, Theorem \ref{thm:BQCADMMdelta} implies $$\left|\frac{1}{n}\log\frac{p_1(y^n)}{p_2(y^n)}\right|<6\rho n\Delta= \frac{1}{n},$$ 
where the last inequality is because $\rho=\frac{1}{12n^2}$.

In the convergent case, we first use a {\emph{necessary}} condition for $x_Q^*=1$ to show that $$\liminf_{n\to\infty}-\frac{1}{n}\log\beta_n\geq C(P_1,P_2).$$ By Theorem \ref{thm:BQCADMMdelta}, when $x_Q^*=1$, $y^n$ must satisfy
$$\left|\mathcal{T_X}\left(\frac{1}{n}\log\frac{p_1(y^n)}{p_2(y^n)}\right)-1\right|<1+4\rho\frac{m}{n}.$$
If we pick $\rho\leq\frac{1}{4m}$ and recall that $\mathcal{T_X}(\cdot)$ is the projection operator that maps a real value to the nearest point in $\mathcal{X}=[-1,1]$, the above inequality indicates that $y^n$ is such that
$$\frac{1}{n}\log\frac{p_1(y^n)}{p_2(y^n)}>-\frac{1}{n}.$$
Together with the cyclic case, we get $$\mathcal{A}_n\subseteq\left\{y^n:\frac{1}{n}\log\frac{p_1(y^n)}{p_2(y^n)}>-\frac{1}{n}\right\}.$$
Hence, $$\beta_n=P_2(\mathcal{A}_n)\leq P_2\left(\left\{y^n:\frac{1}{n}\log\frac{p_1(y^n)}{p_2(y^n)}>- \frac{1}{n}\right\}\right).$$
Comparing with $(\ref{eqn:MAPoptset})$, we see that 
$\{y^n:\frac{1}{n}\log\frac{p_1(y^n)}{p_2(y^n)}>- \frac{1}{n}\}$ is the optimal acceptance region for the hypothesis testing problem with prior probabilities $\frac{2}{3}$ and $\frac{1}{3}$ under $H_1$ and $H_2$, respectively. Therefore, (\ref{eqn:MAPoptset2}) implies that
\begin{align}
\label{eqn:MAPb}
&\liminf_{n\to\infty}-\frac{1}{n}\log\beta_n\nn\\
\geq&\liminf_{n\to\infty}-\frac{1}{n}\log P_2\left(\left\{y^n:\frac{1}{n}\log\frac{p_1(y^n)}{p_2(y^n)}
>-\frac{1}{n}\right\}\right)\nn\\
=&~C(P_1,P_2).
\end{align}

We next find a \emph{sufficient} condition for $y^n\in\mathcal{A}_n$ to establish  
\begin{align}
\label{eqn:MAPa}
\liminf_{n\to\infty}-\frac{1}{n}\log\alpha_n\geq C(P_1,P_2).
\end{align} 
When $y^n\notin\mathcal{A}_n$, convergence must be reached at $x_Q^*=-1$ and we have  
$$\left|\mathcal{T_X}\left(\frac{1}{n}\log\frac{p_1(y^n)}{p_2(y^n)}\right)+1\right|\leq1+4\rho\frac{m}{n}.$$
Therefore, when $n\geq 2$ and $\rho\leq\frac{1}{4m}$, 
$$\mathcal{A}_n^c\subseteq\left\{y^n:\frac{1}{n}\log\frac{p_1(y^n)}{p_2(y^n)}\leq\frac{1}{n}\right\}.$$
Since $\left\{y^n:\frac{1}{n}\log\frac{p_1(y^n)}{p_2(y^n)}\leq\frac{1}{n}\right\}$ is the optimal critical region for the hypothesis testing problem with prior probabilities $\frac{1}{3}$ and $\frac{2}{3}$, (\ref{eqn:MAPa}) can be shown similarly. 

Finally, combining (\ref{eqn:MAPb}) and (\ref{eqn:MAPa}) we have
$$\liminf_{n\to\infty}-\frac{1}{n}\log\left(\pi_1\alpha_n+\pi_2\beta_n\right)\geq\liminf_{n\to\infty}-\frac{1}{n}\log\left(\max\{\pi_1,\pi_2\}\right)+\liminf_{n\to\infty}-\frac{1}{n}\log\left(\max\{\alpha_n,\beta_n\}\right)\geq C(P_1,P_2).$$
The proof is complete by Chernoff theorem and the fact that $m\leq\frac{n(n-1)}{2}$ for a connected undirected graph.
\end{IEEEproof}
\begin{remark}
\label{rmk:rhosel}
It appears that choosing  $\rho=\frac{1}{12n^2}$, which can be very small, may make BQ-CADMM slow. Fortunately, BQ-CADMM is more likely to converge with larger $n$ and we only need $\rho\leq\frac{1}{4m}$ if convergence happens. A decreasing strategy for $\rho$ can also be used to accelerate the convergence of BQ-CADMM; see Section~\ref{sec:SimuD}  as well as \cite[Section~V-D]{Zhu2016BQC}.
\end{remark}

The above theorem indicates that the consensus approach achieves the optimal error exponent which is given by Chernoff theorem. A direct extension is to consider multi-hypothesis testing. We will show that our consensus based approach also achieves the centralized error exponent under MAP criterion by executing multiple runs of BQ-CADMM. Denote the probability measures and  their corresponding prior probabilities  respectively by $P_w$ and $\pi_w$, $w=1,2,\ldots,W$. We also denote $dP_w$ as $p_w$. Assume that all $\pi_w$'s are positive and that $P_w$ and $P_{w'}$ are absolutely mutually continuous with $C(P_w,P_{w'})>0$ for any $w\neq w'$. The centralized MAP rule for the $w$-th hypothesis is given by 
$${A}_w^*(n)=\left\{y^n: \pi_wp_w(y^n)\geq\max_{w'< w}\pi_{w'}p_{w'}(y^n), \pi_wp_w(y^n)>\max_{w'> w}\pi_{w'} p_{w'}(y^n)\right\}.$$
For ease of presentation, define the following Neyman-Pearson test between two different hypotheses $$V_n^{w,w'}=\begin{cases}
\left\{y^n: \frac{1}{n}\log\frac{p_w(y^n)}{p_{w'}(y^n)}\geq \frac{1}{n}\log{\frac{\pi_{w'}}{\pi_{w}}}\right\},~\text{if}~w'<w,\\
\left\{y^n: \frac{1}{n}\log\frac{p_w(y^n)}{p_{w'}(y^n)}>\frac{1}{n}\log{\frac{\pi_{w'}}{\pi_w}}\right\},~\text{if}~w'>w.
\end{cases}$$
Then we can write $A_w^*(n)={\bigcap}_{w'\neq w} V_n^{w,w'}$ for a given $w$. Two useful facts about $V_n^{w,w'}$ are stated as follows:
\begin{itemize}
\item $V_n^{w,w'} = \Sigma^n\setminus V_n^{w',w} \triangleq \left(V_n^{w',w}\right)^c$. 
\item Consider binary hypothesis testing between $P_w$ and $P_w'$ with prior probabilities $\frac{\pi_w}{\pi_w+\pi_{w'}}$ and $\frac{\pi_{w'}}{\pi_w+\pi_{w'}}$, respectively. Then $V_n^{w,w'}$ is the optimal acceptance region for $P_w$ under MAP criterion. Similar to (\ref{eqn:MAPoptset2}), Lemma~\ref{lem:Lambda} and Theorem~\ref{thm:largedev} indicate
$$\liminf_{n\to\infty}-\frac{1}{n}\log P_{w}\left(\left(V_n^{w,w'}\right)^c\right)=\liminf_{n\to\infty}-\frac{1}{n}\log P_{w'}\left(V_n^{w,w'}\right)=C(P_w, P_{w'}).$$
 \end{itemize}
Now consider the optimal Bayesian error for the multi-hypothesis testing
$$P_e^*=\sum_w \pi_w P_w(({A}_w^*(n))^c).$$
Noting that $P_w(({A}_w^*(n))^c)= P_w\left(\bigcup_{w'\neq w}\left(V_n^{w,w'}\right)^c\right)$, we get the following
\begin{align}
\label{eqn:similar}
\max_{w'\neq w}P_w\left(\left(V_n^{w,w'}\right)^c\right)\leq P_w(({A}_w^*(n))^c) \leq \sum_{w'\neq w}P_w\left(\left(V_n^{w,w'}\right)^c\right).
\end{align}
Thus, we get the error exponent $$\liminf_{n\to\infty}-\frac{1}{n}\log P_w(({A}_w^*(n))^c) = \min_{w'\neq w} C(P_w,P_{w'}),$$
for any given $w\in \{1,2,\ldots, W\}$. Hence, the centralized error exponent is $$\liminf_{n\to\infty}-\frac{1}{n}\log P_e^*=\min_w\min_{w'\neq w} C(P_w,P_{w'}).$$

To apply our consensus based approach, we use the bubble sorting idea to construct the MAP detector: 
starting with $w=1$ and $w'=2$, test whether $y^n\in V_n^{w,w'}$; if yes, keep this $w$, otherwise, set $w=w'$; set $w'=w'+1$ and test again if $y^n\in V_n^{w,w'}$; continue this process until the $W$-th hypothesis is involved. It is straightforward to see that the final $w$ is the output of the MAP detector. Recall that the acceptance region in Theorem~\ref{thm:MAP}, when testing between $P_w$ and $P_{w'}$, achieves the same optimal error exponent as $V_n^{w,w'}$. We may replace $V_n^{w,w'}$ with this acceptance region, denoted by $\mathcal{A}_n^{w,w'}$, to implement the consensus based approach. In summary, the above algorithm runs BQ-CADMM $W-1$ times to make a decision for the multi-hypothesis testing problem.

To study how this algorithm performs, let $\mathcal A_w(n)$ be the acceptance region for the $w$-th hypothesis resulting from the $W-1$ runs of BQ-CADMM. For the first hypothesis to be selected, $y^n$ must be such that $y^n\in\bigcap_{w'\neq 1}\mathcal{A}_n^{1,w'}$ and conversely, if $y^n\in\bigcap_{w'\neq 1}\mathcal{A}_n^{1,w'}$, we must select the first hypothesis. Thus, $\mathcal A_1(n)=\bigcap_{w'\neq 1}\mathcal{A}_n^{1,w'}$. Similar to (\ref{eqn:similar}) and using Theorem~\ref{thm:MAP}, we have 
$$\liminf_{n\to\infty} -\frac{1}{n}\log P_1((\mathcal A_1(n))^c) = \min_{w'\neq 1} C(P_1,P_{w'}).$$
For $2\leq w\leq W-1$, if the final decision is $w$, then the $(w-1)$-th run of BQ-CADMM must accept $w$. At the same time, $y^n$ must lie in $\mathcal{A}_n^{w,w'}$ for any $w'>w$. Thus,  $$\left(\bigcap_{w'<w}\mathcal (A_n^{w',w}(n))^c\right)\bigcap\left(\bigcap_{w'>w}\mathcal{A}_n^{w,w'}\right)\subset\mathcal A_w(n)\subset\bigcap_{w'>w}\mathcal{A}_n^{w,w'}.$$
Recalling the definition of $\mathcal A_n^{w,w'}$, we conclude that
$$\min_{w'\neq w}C(P_w,P_{w'})\leq \liminf_{n\to\infty}-\frac{1}{n}\log P_w((\mathcal A_w(n))^c)\leq \min_{w'>w}C(P_w,P_{w'}).$$
For the $W$-th hypothesis, similar argument shows that 
$$\min_{w'\neq W}C(P_W,P_{w'})\leq \liminf_{n\to\infty}-\frac{1}{n}\log P_W((\mathcal A_W(n))^c)\leq \max_{w'\neq W}C(P_W,P_{w'}).$$
Finally, it is noted that the Bayesian error is $P_e=\sum_w \pi_w P_w((\mathcal{A}_w(n))^c)$ and that its error exponent is decided by the lowest error exponent of $P_w((\mathcal{A}_w(n))^c)$. Together with the symmetry property of $C(P_w,P_{w'})$ (i.e., $C(P_w,P_{w'})=C(P_{w'},P_{w})$), we conclude that
\begin{align}
\label{eqn:OptMary}
\liminf_{n\to\infty}-\frac{1}{n}\log P_e=\min_{w}\min_{w'\neq w} C(P_w, P_{w'}),
\end{align}
which is the optimal error exponent in the centralized case. The result is summarized in the following theorem.
\begin{theorem}
Consider multi-hypothesis testing with hypotheses $P_w, w=1,2,\ldots, W$ for some integer $W\geq 2$. Assume that the prior probability $\pi_w$ for each hypothesis is positive and that the hypotheses $P_w$ are pairwise absolutely continuous with each other. Using $W-1$ runs of BQ-CADMM with the same quantizer and algorithm parameters in Theorem~\ref{thm:MAP}, each node can achieve the optimal centralized error exponent under the MAP criterion, which is given in (\ref{eqn:OptMary}).
\end{theorem}

\subsection{Neyman-Pearson Criterion with Exponential Constraint}
\label{sec:NYE}
We now consider the Neyman-Pearson criterion with exponential constraint based on large deviations techniques. Similar to the above two cases, the key is to pick appropriate algorithm and quantizer parameters such that the constructed acceptance region approaches the optimal centralized one as the network size increases.
\begin{theorem}
\label{thm:NYE}
Let $y^n$ be i.i.d.~$\sim Q$. Consider the hypothesis test between $H_1:Q=P_1$ and $H_2:Q=P_2$ and assume that $0<D(P_1\|P_2), D(P_2\|P_1)<\infty$. Set $a=-D(P_2\|P_1)$, $\Delta =D(P_1\|P_2)+D(P_2\|P_1)$, and $\delta = D(P_1\|P_2)+\tau$ with $\tau\in(-D(P_1\|P_2), D(P_2\|P_1))$. Set also $r_i=\log\frac{p_1(y_i)}{p_2(y_i)}$ and $$\rho=\frac{1}{6n^2(D(P_1\|P_2)+D(P_2\|P_1))}.$$ Assume that BQ-CADMM runs sufficiently long such that either convergence or cycling occurs. Let the acceptance region for $H_1$ be
\begin{align}
\mathcal{A}_n=&\big\{y^n\hspace{-1pt}:\text{BQ-CADMM cycles}\}\hspace{2pt}\mathsmaller{\bigcup}\hspace{2pt}\big\{y^n\hspace{-1pt}:\text{BQ-CADMM converges at~}x_Q^*\hspace{-1pt}=\hspace{-1pt}D(P_1\|P_2)\big\}\hspace{-1pt}.\nn
\end{align}
Then we have 
\begin{align}
\liminf_{n\to\infty}-\frac{1}{n}\log\alpha_n = \Lambda^*(\tau),\nn
\end{align}
and 
\begin{align}
\liminf_{n\to\infty}-\frac{1}{n}\log\beta_n = \Lambda^*(\tau)-\tau.\nn
\end{align}
\end{theorem}
\begin{IEEEproof}
The $\delta$-quantizer of this setup has its threshold set at $-\tau$. The proof is similar to previous ones and we hence omit some details.

We first find a \emph{sufficient} condition for $y^n\in\mathcal{A}_n$. If $y^n\not\in\mathcal{A}_n$, BQ-CADMM must reach a consensus at $x_Q^*=-D(P_2\|P_1)$. Then Theorem~\ref{thm:BQCADMMdelta} implies 
\begin{align}
\bigg|\mathcal{T_X}\left(\frac{1}{n}\log\frac{p_1(y^n)}{p_2(y^n)}\right)+D(P_2\|P_1)\bigg|\leq\left(1+4\rho\frac{m}{n}\right)\left(D(P_2\|P_1)-\tau\right).\nn
\end{align}
Picking $\rho$ small enough, e.g., $\rho<\frac{1}{4m(D(P_2\|P_1)-\tau)}$, we get 
\begin{align}
\mathcal{T_X}\left(\frac{1}{n}\log\frac{p_1(y^n)}{p_2(y^n)}\right)\leq-\tau+\frac{1}{n}.\nn
\end{align}
Since $\mathcal {T_X}(\cdot)$ projects a real value into $[-D(P_2\|P_1),D(P_1\|P_2)]$ and $\tau\in(-D(P_1\|P_2), D(P_2\|P_1))$, there exists a positive integer $n_0$ such that for $n\geq n_0$,
$$\mathcal{A}_n^c\subseteq\left\{y^n:\frac{1}{n}\log\frac{p_1(y^n)}{p_2(y^n)}\leq-\tau+\frac{1}{n}\right\}.$$
Therefore, 
\begin{align}
\label{eqn:largedev1}
&\liminf_{n\to\infty}-\frac{1}{n}\log\alpha_n\nn\\
\geq&\liminf_{n\to\infty}-\frac{1}{n}\log P_1\left(\left\{y^n:\frac{1}{n}\log\frac{p_1(y^n)}{p_2(y^n)}\leq -\tau+\frac{1}{n}\right\}\right)\nn\\
=&~\Lambda^*(\tau),
\end{align}
where the last equality is due to Lemma \ref{lem:Lambda} and Theorem~\ref{thm:largedev}.

We next find a {\emph{necessary}} condition for $y^n\in\mathcal{A}_n$. If $y^n$ results in a cyclic behavior of BQ-CADMM, we must have 
$$\left|\frac{1}{n}\log\frac{p_1(y^n)}{p_2(y^n)}+\tau\right|<6\rho n(D(P_1\|P_2)+D(P_2\|P_1)).$$
For the other case where convergence is reached at $x_Q^*=D(P_1\|P_2)$, we have
\begin{align}
\bigg|\mathcal{T_X}\left(\frac{1}{n}\log\frac{p_1(y^n)}{p_2(y^n)}\right)-D(P_1\|P_2)\bigg|<\left(1+4\rho\frac{m}{n}\right)\left(D(P_1\|P_2)+\tau\right).\nn
\end{align}
With $\rho=\frac{1}{6n^2\Delta}$, we can verify that with a sufficiently large $n$,
$$\mathcal{A}_n\subseteq\left\{\frac{1}{n}\log\frac{p_1(y^n)}{p_2(y^n)}>-\tau-\frac{1}{n}\right\}.$$
Thus,
\begin{align}
\label{eqn:largedev2}
\liminf_{n\to\infty}-\frac{1}{n}\log\beta_n\geq\Lambda^*(\tau)-\tau.
\end{align}

Finally, the optimality of the Neyman-Pearson  test and Theorem~\ref{thm:largedev} establish the equalities in (\ref{eqn:largedev1}) and (\ref{eqn:largedev2}). 
\end{IEEEproof}

Therefore, by replacing $\tau=\tau^*$ where $\tau^*$ is the optimal threshold in the centralized case, each node achieves the optimal centralized error exponent in Theorem~\ref{thm:largedev} under the Neyman-Pearson criterion with exponential constraint.
\subsection{Remarks}
We have the following remarks regarding to our main results.

\subsubsection{Parameter Selection}
We focus on the Neyman-Pearson criterion with exponential constraint as both Stein's lemma and Chernoff theorem can be deduced from it. While in Theorem~\ref{thm:NYE} only a single parameter setup is selected for the $\delta$-quantizer to accommodate all possible $-D(P_1\|P_2)<\tau<D(P_2\|P_1)$, the quantizer parameters can be chosen from a quite broad set. With given $\tau$ in the Neyman-Pearson test of (\ref{eqn:NPtest}), one can pick any $a$ and $\Delta$ to satisfy $a<-\tau<a+\Delta$. Then $\delta$ can be chosen such that $0<\delta<\Delta$ and the threshold $ a+\Delta-\delta\to-\tau$ as $n\to\infty$. To guarantee the optimal asymptotic performance, the algorithm parameter $\rho$ must be small enough to ensure $6\rho n\Delta\to 0$ as $n\to\infty$. This setup of quantizer and algorithm parameters can be similarly verified by exploring the sufficient and necessary conditions on the acceptance region. 

It is worth noting that while the optimal error exponent only requires the constructed acceptance region to asymptotically approach the centralized one, prudent choice of  quantizer setup and step size can improve  the non-asymptotic performance; see Section~\ref{sec:nonasymp} and Section~\ref{sec:nonasymp_sim}. Besides, our choice of $\rho$ is obtained from worst-case consensus error bounds that are generally loose and in turn result in loose $\rho$. Thus, our algorithm is very likely to perform well in terms of error probabilities without requiring $\rho$ to be very small; see also the simulation result in Section~\ref{sec:nonasymp_sim} when $n$ is small.
\subsubsection{Practical Considerations}
In our main theorems, the acceptance regions all rely on the average LLR and seem to require knowledge of observations $y^n$ across the sensors. This, however, is not the case; consensus decision making is ensured at each node based only on locally available information through local observation and local information exchange. To see this, assume that either convergence or cycling has occured. Then Theorem~\ref{thm:BQCADMMdelta} guarantees that a consensus is reached at all the nodes. That is, if a node converges at $a$ (or $a+\Delta$), every other node converges at $a$ (or $a+\Delta$); if the node cycles, every other node cycles. As such, each node can make the same decision determined by the acceptance region. 

Since there is no closed-form result on the number of iterations needed for convergence or cycling to happen, there is a need to choose a  stopping criterion at each node.  A natural approach is to set the maximum number of iterations at the beginning, which, however, requires characterization of  the convergence time of BQ-CADMM.  An upper bound on the convergence time, which depends on the network topology and agents' data, may still be insufficient as these quantities are locally unknown. 
In a fully distributed manner, we can run additional algorithms to determine if a consensus has been reached; see, e.g., \cite{Manitara2016,Yadav2007}. These additional algorithms may take a long running time in large networks and will require extra data communications. In general, finding a practically meaningful and efficient stopping criterion for distributed averaging algorithms remains an open problem.  

As to detecting the cyclic state, nodes may record a certain number of consecutive variable values and check if any cycle exists. Yet, there is no guarantee that the oscillation can be perfectly detected, as we do not have a meaningful upper bound on the cyclic period.\footnote{As given in \cite{Zhu2016BQC}, an upper bound can be derived but is too loose to use in practice.} Indeed, the cyclic behavior can be ignored without losing any optimality in terms of error exponents. To see this, consider rejecting $H_1$ if oscillation happens. It can be similarly shown that the same error exponent is achieved under each criterion. Therefore, nodes can make their decision based on their current state to achieve the optimal asymptotic performance, but they fail to reach a consensus when oscillation occurs. Moreover, as shown by our simulation in Section~\ref{sec:nonasymp_sim}, convergence almost always happens with the choice of $\rho=\frac{1}{4m}$ when $n$ becomes large.

\subsubsection{Comparison with Fusion Center Based Structures}
By enabling sensors and fusion center to transmit and receive data, the parallel and tandem networks are equivalent to the undirected star and path graphs, respectively. As such, the parallel and tandem networks can be regarded as special cases of consensus type structures. It is important to note that we achieve the optimal error exponent as in centralized settings at a cost of more data transmissions. The fusion center based structures need $n$ and $n-1$ data transmissions for the parallel and tandem networks, respectively. The consensus based structure, however, has
each sensor sending one bit to its neighbors and hence there are in total $2m$ bits per iteration, where $m=n$ and $n-1$ for the star  and path graphs, respectively. To see how many bits are needed for decision making at all nodes, it requires the characterization on the convergence time of BQ-CADMM. While the current work does not characterize convergence time, we will conduct numerical examples in Section~\ref{sec:simulation} to evaluate it empirically.

\section{Non-asymptotic Detection Performance}
\label{sec:nonasymp}
While this paper focuses on characterizing asymptotic detection performance when the network size scales, we are also interested in non-asymptotic performance of the proposed approach. For a broad class of criteria, including the Neyman-Pearson criterion and the Bayesian criterion, the optimal acceptance region for $H_1$ is defined by a LLR test with a suitably chosen threshold
$$\mathcal{A}_n^*=\left\{y^n:\frac{1}{n}\log\frac{p_1(y^n)}{p_2(y^n)}>\tau^*\right\},$$
where $\tau^*\in\mathbb{R}$ and $n$ is finite. Similar to the asymptotic setting, we will use the consensus result of BQ-CADMM to construct an acceptance region whose type-I and type-II error probabilities are arbitrarily close to the centralized ones.

Set $a = \tau^*-1$, $\Delta=2$, and $\delta =1$ for the $\delta$-quantizer. Then the threshold in this setup is $\tau^*$. Set also $r_i=\log\frac{p_1(y_i)}{p_2(y_i)}$. We again run BQ-CADMM long enough such that either a convergent result or cyclic result is reached. Let the acceptance region for $H_1$ be
\begin{align}
\mathcal{A}_n=&\big\{y^n\hspace{-1pt}:\text{BQ-CADMM cycles}\}~\mathsmaller{\bigcup}~\big\{y^n\hspace{-1pt}:\text{BQ-CADMM converges at~}x_Q^*\hspace{-1pt}=\hspace{-1pt}\tau^*+1\big\}\hspace{-1pt}.\nn
\end{align}

We next find sufficient and necessary conditions for $y^n\in\mathcal{A}_n$. If $y^n\not\in\mathcal{A}_n$, BQ-CADMM converges at $x_Q^*=\tau^*-1$. We get
$$\left|\mathcal{T_X}(\bar{r})-(\tau^*-1)\right|\leq\left(1+4\rho\frac{m}{n}\right).$$ 
Then the projection operator $\mathcal{T_X}(\cdot)$ implies that we must have $y^n\in\mathcal{A}_n$ if $$\bar{r}-\tau^* > 4\rho\frac{m}{n},~\text{if}~\rho<\frac{n}{4m}.$$
When $y^n\in\mathcal{A}_n$, BQ-CADMM either cycles or converges at $x_Q^*=\tau^*+1$. Thus, $y^n$ must be such that either 
$$\left|\bar{r}-\tau^*\right|<12\rho n,$$ or $$\bar{r}-\tau^*>-4\rho\frac{m}{n},~\text{if}~\rho<\frac{n}{4m}.$$
Since $1\leq m\leq \frac{n(n-1)}{2}$ for a connected network, then for $Q\in\{P_1,P_2\}$ and $\rho<\frac{n}{4m}$ we get
\begin{align}
&~Q\left(\left\{y^n:\frac{1}{n}\log\frac{p_1(y^n)}{p_2(y^n)}>\tau^*+4\rho\frac{m}{n}\right\}\right)\nn\\
\leq&~Q(\mathcal{A}_n)\nn\\
\leq&~Q\left(\left\{y^n:\frac{1}{n}\log\frac{p_1(y^n)}{p_2(y^n)}>\tau^*-12\rho n\right\}\right).\nn
\end{align}
Let $z=\frac{1}{n}\log\frac{p_1(y^n)}{p_2(y^n)}$ and denote its cumulative distribution function as $Q(\tau)=Q\left(\left\{z: z\leq\tau\right\}\right)$. We can further write the above as
\begin{align}
1-Q\left(\tau^*+4\rho\frac{m}{n}\right)&\leq Q(\mathcal{A}_n)\leq 1-Q\left(\tau^*-12\rho n\right).\nn
\end{align}
Recall that $Q(\mathcal{A}_n^*)=1-Q(\tau^*),$ we have that $Q(\mathcal{A}_n)\to Q(\mathcal{A}_n^*)$ as $\rho\to 0$ given that $Q(\tau)$ is continuous at $\tau=\tau^*$. Therefore, when the continuity condition holds for both $P_1$ and $P_2$, we can make the type-I and type-II error probabilities arbitrarily close to the optimal ones by picking small enough $\rho$.
\section{Simulations}
\label{sec:simulation}
\subsection{Non-Asymptotic Performance}
\label{sec:nonasymp_sim}
We first consider the following hypothesis testing problem between two Gaussian distributions in star networks:
\begin{itemize}
	\item $H_1: P_1=\mathcal{N}(1,10)$,
	\item $H_2: P_2=\mathcal{N}(-1,10)$.
\end{itemize}
For finite $n$ and positive $\pi_1$ and $\pi_2$, the optimal centralized acceptance region for $H_1$ under the MAP criterion is 
\[\Bigg\{y^n\hspace{-1pt}:\frac{1}{n}\log\frac{p_1(y^n)}{p_2(y^n)}\hspace{-1pt}>\hspace{-1pt}\frac{1}{n}\log\frac{\pi_2}{\pi_1}\Bigg\}\hspace{-1pt}=\hspace{-1pt}\left\{y^n\hspace{-1pt}:\frac{1}5\sum_{i=1}^{n}y_i\hspace{-1pt}>\hspace{-1pt}\ln\frac{\pi_2}{\pi_1}\right\}\hspace{-1pt},\]
and the optimal Bayesian error probability is given by
\begin{align}
\label{eqn:ExampleErr}
P_e^*=&~\pi_1\mathrm{qfunc}\left(\left(1-\frac{5}{n}\ln\frac{\pi_2}{\pi_1}\right)\sqrt{\frac{n}{10}}\right)\nn\\
&~~~~\quad\quad+\pi_2\mathrm{qfunc}\left(\left(1+\frac{5}{n}\ln\frac{\pi_2}{\pi_1}\right)\sqrt{\frac{n}{10}}\right),
\end{align}
where $\mathrm{qfunc}(\cdot)$ denotes the complementary distribution function for standard Gaussian distribution. We perform Monte Carlo simulations to estimate the actual Bayesian error probability of our approach and compare it with (\ref{eqn:ExampleErr}). In the spirit of Remark~\ref{rmk:rhosel}, we run BQ-CADMM with $\rho=\frac{1}{4m}$ and check if BQ-CADMM converges to the right hypothesis. If BQ-CADMM cycles, we rerun BQ-CADMM with $\rho=\frac{1}{12n^2}$ and make the decision based on the new consensus result. Summarized in Fig.~\ref{fig:err} and Fig.~\ref{fig:CT} are the Monte Carlo results for the Gaussian example with different prior probabilities: $\pi_1=0.5$ and $\pi_1=0.1$.

\begin{figure}[!t]
	\centering
	\includegraphics[width=.6\linewidth]{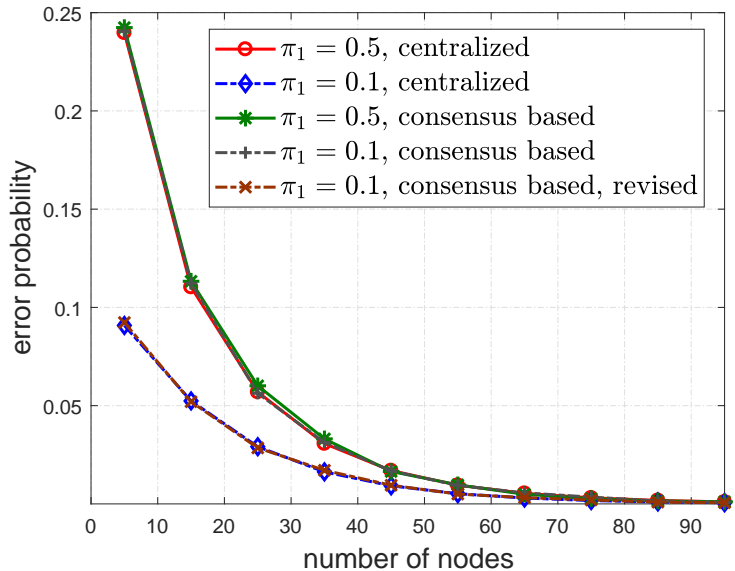}
	\caption{Error probability of Monte Carlo simulations for the Gaussian example; the number of trials for each plotted value is $10^5$.}
	\label{fig:err}
\end{figure}

\begin{figure}[!t]
	\centering
	\includegraphics[width=.6\linewidth]{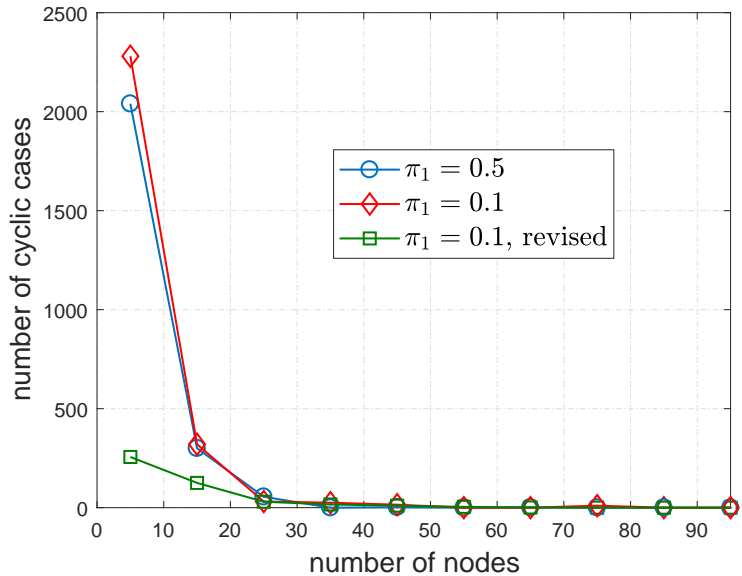}
	\caption{Number of cyclic cases in $10^5$ trials for the Gaussian example.}
	\label{fig:CT}
\end{figure}

We observe from Fig.~\ref{fig:err} that the consensus based error probabilities of the two cases are both very close to the centralized error probability with $\pi_1=0.5$ for all $n$. That said, in the case of $\pi_1=0.1$, the consensus based approach has its error probability far from the centralized one when $n$ is small. This is because we use $\delta=1$ in the consensus based approach for all positive prior probabilities and consequently, the acceptance region is very different from the optimal centralized one with $\pi_1=0.1$ for small $n$. Similar to the setup of the $\delta$-quantizer in Section~\ref{sec:nonasymp}, we pick $\delta=1-\frac{1}{n}\log\frac{\pi_2}{\pi_1}$ such that the threshold of the $\delta$-quantizer becomes $\frac{1}{n}\log\frac{\pi_2}{\pi_1}$.  Here we need $n\geq 4$ to ensure  $\delta>0$.  Running the example with $\pi_1=0.1$ again, we obtain the new error probability very close to the optimal one (see the dashed line with cross markers in Fig.~\ref{fig:err}). In addition, we record the number of trials in which BQ-CADMM cycles with $\rho=\frac{1}{4m}$. As plotted in Fig.~\ref{fig:CT}, it is clear that BQ-CADMM tends to converge as $n$ increases. 

\subsection{Convergence Time}
\label{sec:SimuD}
We now numerically evaluate convergence time of the proposed approach. Given the number of nodes, we consider the above Gaussian example with $\pi_1=0.5$ over star graph which has the smallest number of edges for a connected network, complete graph which has the largest number of edges, and randomly generated connected graphs with intermediate numbers of edges. A random graph with $n$ nodes and $m$ edges is generated as follows: first generate a complete graph of $n$ nodes and then randomly remove
$\frac{n(n-1)}{2}-m$ edges while ensuring the graph stays connected. Since we have shown that BQ-CADMM with this example is more likely to converge with larger $n$, we only count the convergent cases and pick $\rho=\frac{1}{4m}$ for BQ-CADMM. 

Simulation result is shown in Fig.~\ref{fig:ACT}. {The plotted value is the average of $2,000$ runs in which both data and graph are randomly generated at each run.} One can see that sparser and larger networks usually have longer convergence time and that the average convergence time for all cases is approximately $O(n\log n)$. As a result, in a sensor network with $n$ nodes and $m$ edges, the proposed approach requires approximately $O(mn\log n)$ bits of data transmissions for the Gaussian example.
\begin{figure}[htp]
\centering
    \includegraphics[width=0.6\linewidth]{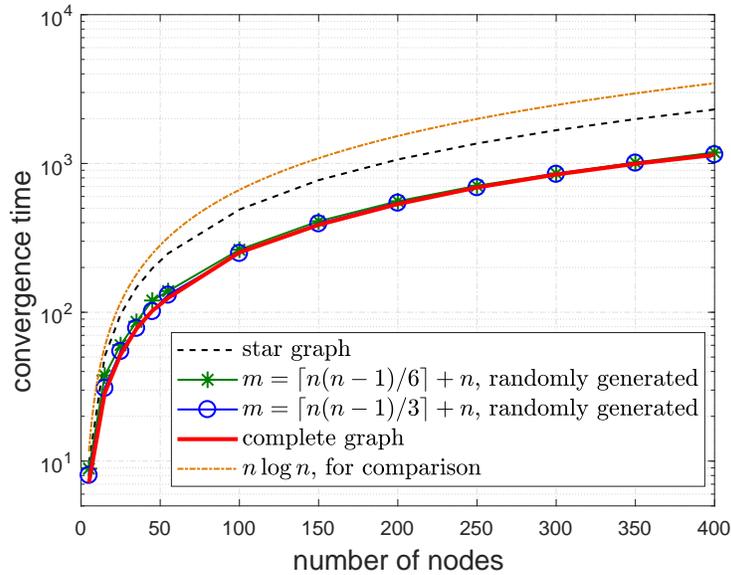}
      \caption{Convergence time of BQ-CADMM for the Gaussian example with $\pi_1=0.5$; each plotted value is the average of $2,000$ runs.}
      \label{fig:ACT}
\end{figure}

Noticing that the above simulation uses a fixed algorithm parameter for BQ-CADMM, we now apply a decreasing parameter strategy  which is shown to dramatically reduce the convergence time in \cite{Zhu2016BQC}. Start with $\rho=\frac{n}{m}$. If $\rho >\frac{1}{4m}$, we run BQ-CADMM for $50$ iterations and then reduce $\rho$ by a factor of $10$. We repeat this process until $\rho\leq\frac{1}{4m}$ at which we run BQ-CADMM long enough such that either convergence or cycling occurs. The average convergence time is shown in Fig.~\ref{fig:ACT_dec}. Compared with fixed parameter strategy, we observe that the decreasing strategy runs $50\left\lceil\log_{10}(4n)\right\rceil$ iterations before $\rho$ meets the accuracy requirement and makes BQ-CADMM proceed faster at early stages. When $\rho$ indeed satisfies the required accuracy, it only takes a few iterations before reaching the final state. With the decreasing parameter strategy for BQ-CADMM, we conjecture that the consensus based approach requires $O(m\log n)$ bits of data transmissions on the average to achieve the optimal asymptotic performance for the Gaussian example.

\begin{figure}[htp]
\centering
    \includegraphics[width=.6\linewidth]{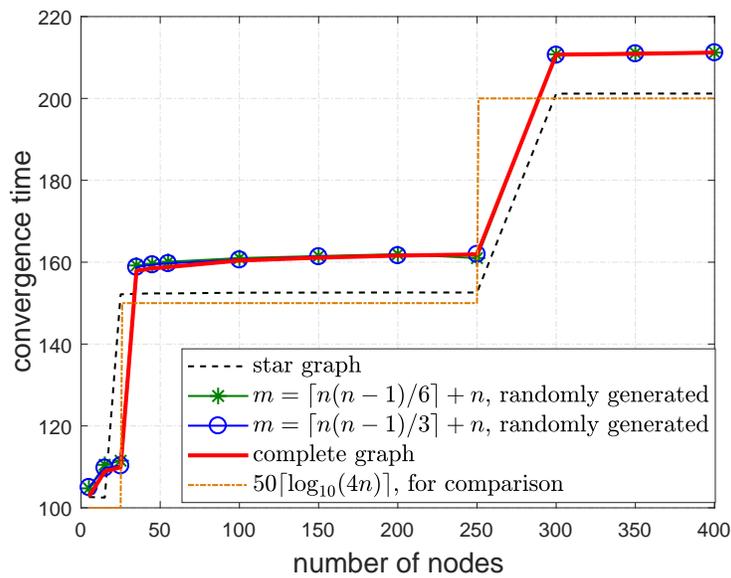}
      \caption{Convergence time of BQ-CADMM using decreasing parameter strategy for the Gaussian example with $\pi_1=0.5$; each plotted value is the average of $2,000$ runs.}
      \label{fig:ACT_dec}
\end{figure}

\section{Conclusion and Discussion}
\label{sec:conclusion}
This paper studies asymptotic performance of consensus based detection for large scale connected networks using BQ-CADMM, a recently proposed quantized consensus approach that has finite quantization levels with possibly unbounded data. Different from the original BQ-CADMM that has a constant term in the consensus error bound, we construct a binary quantizer with controllable threshold such that the consensus value can be of desired accuracy. We then show that each node can achieve the same optimal error exponent as the centralized cases under three common criteria. Non-asymptotic behavior of the proposed approach is also addressed.

While the convergence time of BQ-CADMM is shown to be finite, extensive numerical studies show that  the decreasing strategy in our simulations requires approximately $O(\log n)$ iterations for the proposed approach to reach a decision. However, precise characterization on convergence time or its upper bound remains elusive. Meanwhile, many works, such as \cite{Kar2013,Jakovetic2012,Bajovic2012,Braca2010}, consider the sequential setting where the total number of sensors is finite and each sensor can receive successive observations. Our future work would be to extend our approach, with modifications to handle temporal observations, to the sequential setting. 

\appendix
\begin{IEEEproof}[Proof of Theorem~\ref{thm:BQCADMMdelta}]By the definition of $\mathcal{Q}_\delta(\cdot)$ in (\ref{eqn:detalQDef}), it is clear that $\left|\mathcal{Q}_\delta(x)-\mathcal{T_X}({x})\right|<\Delta<\infty$ for $x\in\mathbb{R}$. Together with the deterministic property (i.e., $\mathcal Q_\delta(x_1)=\mathcal{Q}_\delta(x_2)$ if $x_1=x_2$), that BQ-CADMM using this $\delta$-quantizer either converges or cycles with every node having the same sum of quantized variable values over one period can be shown by the same idea as that of \cite[Theorem 3]{Zhu2016BQC}. What remains is to derive the error bounds in the respective cases.

{\it Convergent case:} With convergence, we can write $x_i^k=x_i^*$ for $k\geq k_0$. To show the error bound (\ref{eqn:bounddQ}), denote $e_i^*=\mathcal{T_X}(x_i^*)-x_Q^*$.  Consider first $x_Q^*=a$ and assume that $|x_Q^*-\mathcal{T}_\mathcal{X}(\bar{r})|>\Delta-\delta$; otherwise, the error bound holds trivially. Following the same steps of the consensus error proof of \cite[Theorem 3]{Zhu2016BQC}, we obtain that
\begin{align}
\left|x_Q^*-\mathcal{T_X}(\bar{r})\right|\leq\frac{1}{n}\sum_{i=1}^n\left(2\rho|\mathcal{N}_i|+1\right)|e_i^*|.\nn
\end{align}
Since $\sum_{i=1}^n|\mathcal{N}_i|=2m$, it remains to find an upper bound for $e_i^*$. Note that $\mathcal{Q}_\delta(x_i^*)=x_Q^*=a$ implies $\mathcal{T_X}(x_i^*)\in[a,a+\Delta-\delta]$ and hence $|e_i^*|=|\mathcal{T_X}(x_i^*)-a|\leq\Delta-\delta$. Similar argument proves the error bound for $x_Q^*=a+\Delta$ and is omitted.

{\it Cyclic case:} We will show that $x_i^k$ is close to the threshold $a+\Delta-\delta$ and then use this fact to characterize the difference between the threshold and the data average. Assume that the cycling state has been reached, i.e., $k\geq k_0$. With only two quantization values $a$ and $a+\Delta$, there are at most four possible cases for two consecutive agent values $\mathcal Q_\delta(x_i^k)$ and $\mathcal{Q}_\delta(x_i^{k+1})$. We now discuss these cases one by one and before this, we write the following useful update from the $x_i$- and $\alpha_i$-updates in Algorithm~1:
\begin{align}
\label{eqn:useful}
x_i^{k+1}&=\frac{1}{1+2\rho|\mathcal{N}_i|}\Bigg(\rho|\mathcal{N}_i|\mathcal Q_\delta(x_i^{k})+\rho\sum_{j\in\mathcal{N}_i}\mathcal Q_\delta(x_j^{k})-\alpha_i^{k}+r_i\Bigg)\nn\\
&=\frac{1}{1+2\rho|\mathcal{N}_i|}\Bigg(\rho|\mathcal{N}_i|\mathcal Q_\delta(x_i^{k})+\rho\sum_{j\in\mathcal{N}_i}\mathcal Q_\delta(x_j^{k})-\alpha_i^{k-1}-\rho|\mathcal{N}_i|\mathcal Q_\delta(x_i^k)+\rho\sum_{j\in\mathcal{N}_i}\mathcal Q_\delta(x_j^k)+r_i\Bigg)\nn\\
&=x_i^{k}+\frac{1}{1+2\rho|\mathcal{N}_i|}\Bigg(2\rho\hspace{-3pt}\sum_{j\in\mathcal{N}_i}\mathcal Q_\delta(x_j^{k})-\rho|\mathcal{N}_i|\mathcal Q_\delta(x_i^{k-1})-\rho\sum_{j\in\mathcal{N}_i}\mathcal Q_\delta(x_j^{k-1})\Bigg).
\end{align}
\begin{itemize}
\item Case 1:~$\mathcal Q_\delta(x_i^k)=a$ and $\mathcal{Q}_\delta(x_i^{k+1})=a+\Delta$. By the definition of $\delta$-quantizer, we have $x_i^{k}\leq a+\Delta-\delta<x_i^{k+1}$. Following (\ref{eqn:useful}) and using the fact that only $a$ and $a+\Delta$ can be the output of $\mathcal{Q}_\delta(\cdot)$, we also get $x_i^{k+1}\leq a+\Delta-\delta+\frac{2\rho|\mathcal{N}_i|}{1+2\rho|\mathcal{N}_i|}\Delta$. In summary, we have 
$$a+\Delta-\delta < x_i^{k+1}\leq a+\Delta-\delta+\frac{2\rho|\mathcal{N}_i|}{1+2\rho|\mathcal{N}_i|}\Delta.$$
\item Case 2:~$\mathcal Q_\delta(x_i^k)=a+\Delta$ and $\mathcal{Q}_\delta(x_i^{k+1})=a$. Similar to Case~1, it can be shown that
$$a+\Delta-\delta -\frac{2\rho|\mathcal{N}_i|}{1+2\rho|\mathcal{N}_i|}\Delta< x_i^{k+1}\leq a+\Delta-\delta.$$
 \item Case 3:~$\mathcal Q_\delta(x_i^k)=a+\Delta$ and $\mathcal{Q}_\delta(x_i^{k+1})=a+\Delta$. We can immediately conclude that $x_i^{k+1}>a+\Delta-\delta$. To find an upper bound on $x_i^{k+1}$, consider the $\alpha_i$-update at index $k$: \begin{align}
\alpha_i^{k}&=\alpha_i^{k-1}+\rho|\mathcal{N}_i|\mathcal Q_\delta(x_i^{k})-\rho\sum_{j\in\mathcal{N}_i}\mathcal Q_\delta(x_j^{k})\geq\alpha_i^{k-1},\nn
\end{align}
where inequality follows from $\mathcal{Q}_\delta(x_i^k)=a+\Delta$. By induction, we have $\alpha_i^{k'}\leq\alpha_i^{k}$ where ${k'}< k$ is the largest index such that $\mathcal{Q}_\delta(x_i^{k'})=a$. Note that such $k'$ always exists for $k\geq k_0+T$ as a result of the cyclic behavior and (\ref{eqn:cyceqn}). Then we have   
\begin{align}
x_i^{k+1}=&~\frac{1}{1+2\rho|\mathcal{N}_i|}\Bigg(\rho|\mathcal{N}_i|\mathcal Q_\delta(x_i^{k})+\rho\sum_{j\in\mathcal{N}_i}\mathcal Q_\delta(x_j^{k})-\alpha_i^{k}+r_i\Bigg)\nn\\
\leq&~\frac{1}{1+2\rho|\mathcal{N}_i|}\Bigg(\rho|\mathcal{N}_i|\mathcal Q_\delta(x_i^{k})+\rho\sum_{j\in\mathcal{N}_i}\mathcal Q_\delta(x_j^{k})-\alpha_i^{k'}+r_i\Bigg)\nn\\
=&~x_i^{k'}+ \frac{1}{1+2\rho|\mathcal{N}_i|}\Bigg(\rho|\mathcal{N}_i|\mathcal Q_\delta(x_i^{k})+\rho\sum_{j\in\mathcal{N}_i}\mathcal Q_\delta(x_j^{k})+\rho\sum_{j\in\mathcal{N}_i}\mathcal Q_\delta(x_j^{k'})-\rho|\mathcal{N}_i|\mathcal Q_\delta(x_i^{k'})\nn\\ 
&-\rho|\mathcal{N}_i|\mathcal Q_\delta(x_i^{k'-1})-\rho\sum_{j\in\mathcal{N}_i}\mathcal Q_\delta(x_j^{k'-1})\Bigg),\nn
\end{align}
where the last equality is obtained by the $x_i$- and $\alpha_i$-updates at the $k'$-th iteration. As $\mathcal{Q}_\delta(x_i^{k'})=a$,  we have $$x_i^{k+1}\leq a+\Delta-\delta+\frac{3\rho|\mathcal{N}_i|}{1+2\rho|\mathcal{N}_i|}\Delta.$$
\item Case~4: $\mathcal Q_\delta(x_i^k)=a$ and $\mathcal{Q}_\delta(x_i^{k+1})=a$. Similar to Case~3, we can get 
$$a+\Delta-\delta -\frac{3\rho|\mathcal{N}_i|}{1+2\rho|\mathcal{N}_i|}\Delta< x_i^{k+1}\leq a+\Delta-\delta.$$
\end{itemize}
Summarizing all the four cases and using the fact that $|\mathcal{N}_i|<n$, we conclude the following:
\begin{align}
\label{eqn:xbdcyc}
|x_i^{k}-(a+\Delta-\delta)|\leq\frac{3\rho|\mathcal{N}_i|}{1+2\rho|\mathcal{N}_i|}\Delta<\frac{3\rho n}{1+2\rho n}\Delta,~\text{for}~k\geq k_0.
\end{align}

With (\ref{eqn:xbdcyc}), we can now bound $x_i^k-\mathcal{Q}_\delta(x_i^k)$ with $k\geq k_0$. If $\mathcal{Q}_\delta(x_i^k)=a$, then $$a+\Delta-\delta-\frac{3\rho n}{1+2\rho n}\Delta< x_i^k\leq a+\Delta-\delta.$$ Thus, $$|x_i^k-\mathcal{Q}_\delta(x_i^k)|\leq\max\left\{\left|\Delta-\delta-\frac{3\rho n}{1+2\rho n}\Delta\right|,\Delta-\delta\right\}.$$ When $\mathcal{Q}_\delta(x_i^k)=a+\Delta$, we can similarly derive  
\begin{align}
|x_i^k-\mathcal{Q}_\delta(x_i^k)|\leq\max\left\{\left|\frac{3\rho n}{1+2\rho n}\Delta-\delta\right|,\delta\right\}.\nn
\end{align}
As $0<\delta<\Delta$, we finally have 
\begin{align}
\label{eqn:bdbddd}
|x_i^k-\mathcal{Q}_\delta(x_i^k)|< \frac{3}{2}\Delta.
\end{align}

To bound the difference between the threshold $a+\Delta-\delta$ and the data average $\bar{r}$, we sum up the variable values over one period for $k\geq k_0$ and get  
\begin{align}
\label{eqn:eqncyc2}
\left|\frac{\sum_{l=0}^{T-1} x_i^{k+l}}{T}-(a+\Delta-\delta)\right|<\frac{3\rho n}{1+2\rho n}\Delta,
\end{align}
and 
\begin{align}
\label{eqn:eqncyc3}
\left|\frac{\sum_{l=0}^{T-1} x_i^{k+l}}{T}-\frac{\sum_{l=0}^{T-1}\mathcal{Q}_\delta(x_i^{k+l})}{T}\right| < \frac{3}{2}\Delta,
\end{align}
where (\ref{eqn:eqncyc2}) is from (\ref{eqn:xbdcyc}) and (\ref{eqn:eqncyc3}) is from (\ref{eqn:bdbddd}).
Also summing up both sides of the $x_i$-update of BQ-CADMM over a period yields
\begin{align}
\label{eqn:sss}
(1+2\rho|\mathcal{N}_i|)\frac{\sum_{l=0}^{T-1}x_i^{k+l}}{T}-\rho|\mathcal{N}_i|\frac{\sum_{l=0}^{T-1}\mathcal{Q}_\delta(x_i^{k+l})}{T}-\rho\sum_{j\in\mathcal{N}_i}\frac{\sum_{l=0}^{T-1}\mathcal{Q}_\delta(x_j^{k+l})}{T}+\frac{\sum_{l=0}^{T-1} \alpha_i^{k+l}}{T}-r_i=0.
\end{align}
Further summing up both sides (\ref{eqn:sss}) from $i=1$ to $n$, we have
\begin{align}
\label{eqn:sss2}
\sum_{i=1}^n(1+2&\rho|\mathcal{N}_i|)\frac{\sum_{l=0}^{T-1}x_i^{k+l}}{T}-\sum_{i=1}^{n}2\rho|\mathcal{N}_i|\frac{\sum_{l=0}^{T-1}\mathcal{Q}_\delta(x_i^{k+l})}{T}
-\sum_{i=1}^nr_i=0,
\end{align}
where we use the fact $\sum_{i=1}^n \alpha_i^k=0$ for any $k$ (cf. \cite[Lemma~1]{Zhu2016BQC}) together with (\ref{eqn:cyceqn}). To complete the proof, we divide both sides of (\ref{eqn:sss2}) by $n$ and use (\ref{eqn:eqncyc2}) and (\ref{eqn:eqncyc3}), which leads to
\begin{align}
\left|\bar{r}-(a+\Delta-\delta)\right|&< 4\rho\frac{m}{n}\frac{3}{2}\Delta+\frac{3\rho n}{1+2\rho n}\Delta<6\rho n \Delta.\nn
\end{align}
where the second inequality is because $m\leq\frac{n(n-1)}{2}$ for a connected undirected graph.
\end{IEEEproof}

\bibliographystyle{IEEEtran}
\bibliography{SZhuBib}

\end{document}

%% file: macro_jnl.tex
\newtheorem{theorem}{Theorem}
\newtheorem{example}{Example}

\newtheorem{lemma}{Lemma}
\newtheorem{definition}{Definition}

\def\psfancypar#1#2{\begingroup\def\par{\endgraf\endgroup\lineskiplimit=0pt}
               \setbox2=\hbox{\large\sc #2}
               \newdimen\tmpht \tmpht \ht2 \advance\tmpht by \baselineskip
               \font\hhuge=Times-Bold at \tmpht
               \setbox1=\hbox{{\hhuge #1}}
               \count7=\tmpht \count8=\ht1
               \divide\count8 by 1000 \divide\count7 by \count8
               \tmpht=.001\tmpht\multiply\tmpht by \count7
               \font\hhuge=Times-Bold at \tmpht
               \setbox1=\hbox{{\hhuge #1}}
               \noindent
                \hangindent1.05\wd1
               \hangafter=-2 {\hskip-\hangindent
               \lower1\ht1\hbox{\raise1.0\ht2\copy1}%
                \kern-0\wd1}\copy2\lineskiplimit=-1000pt}

\newcommand{\beq}{\begin{equation}}
\newcommand{\eeq}{\end{equation}}
\newcommand{\bqa}{\begin{eqnarray}}
\newcommand{\eqa}{\end{eqnarray}}
\newcommand{\bqn}{\begin{eqnarray*}}
\newcommand{\eqn}{\end{eqnarray*}}
\newcommand{\nn}{\nonumber}

\newcommand{\be}{\begin{enumerate}}
\newcommand{\ee}{\end{enumerate}}
\newcommand{\bi}{\begin{itemize}}
\newcommand{\ei}{\end{itemize}}
\newcommand{\bd}{\begin{description}}
\newcommand{\ed}{\end{description}}
\newcommand{\ba}{\begin{array}}
\newcommand{\ea}{\end{array}}
\newcommand{\bde}{\begin{definition}}
\newcommand{\ede}{\end{definition}}
\newcommand{\bex}{\begin{example}}
\newcommand{\eex}{\end{example}}

\def\boxit#1{\vbox{\hrule\hbox{\vrule\kern3pt
        \vbox{\kern3pt#1\kern3pt}\kern3pt\vrule}\hrule}}

\def\reals{ { {\rm  I \kern-0.15em R }  } }
\def\complex{ {\,{{\rm C} \kern-0.50em \raise0.20ex {  |}}\, }}

\def\0bf{{\bf 0}}
\def\1bf{{\bf 1}}
\def\2bf{{\bf 2}}
\def\3bf{{\bf 3}}
\def\4bf{{\bf 4}}
\def\5bf{{\bf 5}}
\def\6bf{{\bf 6}}
\def\7bf{{\bf 7}}
\def\8bf{{\bf 8}}
\def\9bf{{\bf 9}}

\def\Rbf{{\bf R}}

\def\Rxx{\Rbf_{\ssstyle X\kern-.1em X}}

\let\ssstyle=\scriptscriptstyle

\def\Kout{\setbox1=\hbox{\Huge\bf K}\hbox to
1.05\wd1{\hspace{.05\wd1}
}}
\def\Sout{\setbox1=\hbox{\Huge\bf S}\hbox to 1.05\wd1{\hspace{.05\wd1}
}}

%% file: DistDetection_consensus_Part2_onecol.bbl
\begin{thebibliography}{10}
\providecommand{\url}[1]{#1}
\csname url@samestyle\endcsname
\providecommand{\newblock}{\relax}
\providecommand{\bibinfo}[2]{#2}
\providecommand{\BIBentrySTDinterwordspacing}{\spaceskip=0pt\relax}
\providecommand{\BIBentryALTinterwordstretchfactor}{4}
\providecommand{\BIBentryALTinterwordspacing}{\spaceskip=\fontdimen2\font plus
\BIBentryALTinterwordstretchfactor\fontdimen3\font minus
  \fontdimen4\font\relax}
\providecommand{\BIBforeignlanguage}[2]{{%
\expandafter\ifx\csname l@#1\endcsname\relax
\typeout{** WARNING: IEEEtran.bst: No hyphenation pattern has been}%
\typeout{** loaded for the language `#1'. Using the pattern for}%
\typeout{** the default language instead.}%
\else
\language=\csname l@#1\endcsname
\fi
#2}}
\providecommand{\BIBdecl}{\relax}
\BIBdecl

\bibitem{Zhu2016ISIT}
S.~Zhu and B.~Chen, ``Distributed detection over connected networks via one-bit
  quantizer,'' in \emph{Proc. IEEE Int. Symp. Inf. Theory}, Barcelona, Spain,
  Jul. 2016.

\bibitem{Varshney1997distributed}
P.~K. Varshney, \emph{Distributed Detection and Data Fusion}.\hskip 1em plus
  0.5em minus 0.4em\relax New York: Springer, 1997.

\bibitem{Viswanathan1997}
R.~Viswanathan and P.~K. Varshney, ``Distributed detection with multiple
  sensors: {Part I---Fundamentals},'' \emph{Proc. IEEE}, vol.~85, no.~1, pp.
  54--63, 1997.

\bibitem{Blum1997}
R.~S. Blum, S.~A. Kassam, and H.~V. Poor, ``Distributed detection with multiple
  sensors: {Part II---Advanced topics},'' \emph{Proc. IEEE}, vol.~85, no.~1,
  pp. 64--79, 1997.

\bibitem{Chamberland2003}
J.~F. \vspace{0mm}Chamberland and V.~V. Veeravalli, ``Decentralized detection
  in sensor networks,'' \emph{IEEE Trans. Signal Process.}, vol.~51, no.~2, pp.
  407--416, 2003.

\bibitem{Chamberland2004}
J.~F. Chamberland and V.~V. Veeravalli, ``Asymptotic results for decentralized
  detection in power constrained wireless sensor networks,'' \emph{IEEE J. Sel.
  Areas Commun.}, vol.~22, no.~6, pp. 1007--1015, Aug. 2004.

\bibitem{Tsitsiklis1988}
J.~N. Tsitsiklis, ``Decentralized detection by a large number of sensors,''
  \emph{Mathematics of Control, Signals and Systems}, vol.~1, no.~2, pp.
  167--182, 1988.

\bibitem{Tay2008}
W.~P. Tay, J.~N. Tsitsiklis, and M.~Z. Win, ``On the subexponential decay of
  detection error probabilities in long tandems,'' \emph{IEEE Trans. Inf.
  Theory}, vol.~54, no.~10, pp. 4767--4771, 2008.

\bibitem{Saligrama2006}
V.~Saligrama, M.~Alanyali, and O.~Savas, ``Distributed detection in sensor
  networks with packet losses and finite capacity links,'' \emph{IEEE Trans.
  Signal Process.}, vol.~54, no.~11, pp. 4118--4132, Nov. 2006.

\bibitem{Kar2008}
S.~Kar, S.~Aldosari, and J.~M.~F. Moura, ``Topology for distributed inference
  on graphs,'' \emph{IEEE Trans. Signal Process.}, vol.~56, no.~6, pp.
  2609--2613, Jun. 2008.

\bibitem{Kar2011}
S.~Kar, R.~Tandon, H.~V. Poor, and S.~Cui, ``Distributed detection in noisy
  sensor networks,'' in \emph{Proc. IEEE Int. Symp. Inf. Theory}, Saint
  Petersburg, Russia, Jul. 2011.

\bibitem{Kar2013}
S.~Kar and J.~M.~F. Moura, ``Consensus + innovations distributed inference over
  networks: cooperation and sensing in networked systems,'' \emph{IEEE Signal.
  Proc. Mag.}, vol.~30, no.~3, pp. 99--109, May 2013.

\bibitem{Jakovetic2012}
D.~Jakovetic, J.~M.~F. Moura, and J.~Xavier, ``Distributed detection over noisy
  networks: Large deviations analysis,'' \emph{IEEE Trans. Signal Process.},
  vol.~60, no.~8, pp. 4306--4320, Aug. 2012.

\bibitem{Bajovic2012}
D.~Bajovic, D.~Jakovetic, J.~M.~F. Moura, J.~Xavier, and B.~Sinopoli, ``Large
  deviations performance of consensus+innovations distributed detection with
  non-gaussian observations,'' \emph{IEEE Trans. Signal Process.}, vol.~60,
  no.~11, pp. 5987--6002, Nov. 2012.

\bibitem{Braca2010}
P.~Braca, S.~Marano, V.~Matta, and P.~Willett, ``Asymptotic optimality of
  running consensus in testing binary hypotheses,'' \emph{IEEE Trans. Signal
  Process.}, vol.~58, no.~2, pp. 814--825, Feb. 2010.

\bibitem{Kashyap2007}
A.~Kashyap, T.~Ba{\c{s}}ar, and R.~Srikant, ``Quantized consensus,''
  \emph{Automatica}, vol.~43, no.~7, pp. 1192--1203, 2007.

\bibitem{Nedic2009}
A.~Nedic, A.~Olshevsky, A.~Ozdaglar, and J.~N. Tsitsiklis, ``On distributed
  averaging algorithms and quantization effects,'' \emph{IEEE Trans. Autom.
  Control}, vol.~54, no.~11, pp. 2506--2517, Nov. 2009.

\bibitem{Kar2010}
S.~Kar and J.~M. Moura, ``Distributed consensus algorithms in sensor networks:
  quantized data and random link failures,'' \emph{IEEE Trans. Signal
  Process.}, vol.~58, no.~3, pp. 1383--1400, 2010.

\bibitem{Carli2010}
R.~Carli, F.~Fagnani, P.~Frasca, and S.~Zampieri, ``Gossip consensus algorithms
  via quantized communication,'' \emph{Automatica}, vol.~46, no.~1, pp. 70--80,
  2010.

\bibitem{Zhu2016TSP}
S.~Zhu and B.~Chen, ``Quantized cosensus by the {ADMM}: probabilistic versus
  deterministic quantizers,'' \emph{IEEE Trans. Signal Process.}, vol.~64,
  no.~7, pp. 1700--1713, Apr. 2016.

\bibitem{Li2011}
T.~Li, M.~Fu, L.~Xie, and J.~F. Zhang, ``Distributed consensus with limited
  communication data rate,'' \emph{IEEE Trans. Autom. Control}, vol.~56, no.~2,
  pp. 279--292, Feb. 2011.

\bibitem{Zhu2016BQC}
S.~Zhu and B.~Chen, ``Distributed average consensus with bounded quantizer and
  unbounded input,'' \emph{arXiv preprint arXiv:1602.04193}, 2016.

\bibitem{Dembo2009}
A.~Dembo and O.~Zeitouni, \emph{Large Deviations Techniques and
  Applications}.\hskip 1em plus 0.5em minus 0.4em\relax New York: Springer,
  2009.

\bibitem{Cover2006}
T.~M. Cover and J.~A. Thomas, \emph{Elements of Information Theory},
  2nd~ed.\hskip 1em plus 0.5em minus 0.4em\relax New York: Wiley, 2006.

\bibitem{Hoeffding1965}
W.~Hoeffding, ``Asymptotically optimal tests for multinomial distributions,''
  \emph{The Annals of Mathematical Statistics}, pp. 369--401, 1965.

\bibitem{Manitara2016}
N.~E. Manitara and C.~N. Hadjicostis, ``Distributed stopping for average
  consensus in undirected graphs via event-triggered strategies,''
  \emph{Automatica}, vol.~70, pp. 121--127, 2016.

\bibitem{Yadav2007}
V.~Yadav and M.~V. Salapaka, ``Distributed protocol for determining when
  averaging consensus is reached,'' in \emph{45th Annual Allerton Conference on
  Communication, Control, and Computing}, Monticello, IL, 2007.

\end{thebibliography}
